# Multimodal Machine Learning for 3-Dimensional Characterization of Hidden Groundwater and Geothermal Resources

Michael J. Friedel[1,2,3], Nicole Lautze[1], Erin Wallin[1], Roland Gritto[4], Alain Bonneville[5], Massimo Buscema[3], Steve Martel[1]

[1] Hawai'i Groundwater & Geothermal Resource Center, University of Hawai'i-Manoa, USA

[2] Department of Physics, University of Colorado-Denver, USA

[3] Semeion Institute, Rome, Italy

[4] EMR Solutions & Technology, Berkeley, CA, USA

[5] Pacific Northwest National Laboratory, WA, USA

mfriedel@hawaii.edu

## ABSTRACT

The availability of freshwater and low-cost electricity are two limiting factors for sustainable living in Hawai'i and worldwide. This fact raises the question: Can technology be developed to locate freshwater and geothermal resources simultaneously? We present a multimodal machine learning (MML) workflow to assimilate and simultaneously predict the 3d distribution of numeric and categorical features (field observations) along a groundwater-geothermal continuum. Success of the MML workflow relies on a transductive learning algorithm that projects field modalities onto a single embedding space (hypersurface). Multimodalities can include any combination of *measured* (point field) and *derived* (multiphysics-based numerical model inversions, data-driven machine learning, and multiphysics-informed machine learning) *features.* The proposed MML workflow is applied to assimilate randomly shuffled subsets of Hawai'i Play Fairway modalities and predict subsurface geophysical, geologic, and hydrogeologic features at the Islands of Lānaʻi and Hawai'i. Despite challenging field data characteristics (disparate, scale dependent, spatially limited, sparse, and uncertain), the MML workflow yields a single 3d transdisciplinary model that generalizes well to independent data presented to the trained model. The predicted features are used to identify hidden groundwater and geothermal resources at Lānaʻi, and geothermal resources at Hawai'i. Other interpreted subsurface features at Lānaʻi include basalt, batholith, dike swarm, pluton, sill, mantle, Moho, and 3d geothermal stratigraphic units; whereas interpreted subsurface features at Hawai'i include 3d velocity layering, 3d earthquake-fault associations, 3d fault systems; basalt, oceanic crust, magmatic underplating, lithospheric flexure, mantle, and Moho. This study provides new capabilities for characterizing continuous subsurface groundwater and geothermal features for sustainable living in the Hawai'ian Islands and other geothermal sites worldwide.

## 1. INTRODUCTION

The availability of freshwater and low-cost electricity are two important factors for sustainable living in Hawai'i and worldwide. This raises the question: *Can technology be developed to locate and characterize freshwater and geothermal resources simultaneously*? The Hawai'i Play Fairway project, funded by the Geothermal Technologies Office (GTO), U.S. Department of Energy (DOE), produced a statistical methodology to integrate existing geologic, groundwater, and geophysical datasets relevant to subsurface heat, fluid, and permeability into a statewide resource probability map (Lautze et al., 2020). The combined probabilities for heat, fluid, and permeability indicated the likelihood of a geothermal resource in the caldera region on the Island of Lānaʻi Island and the Island of Hawai'i. The prospective geothermal resource area on Lānaʻi, known as the Pālāwai Basin, and on Hawai'i, the Lower East Rift Zone (LERZ), exhibits co-located high gravity signature, reduced resistivities, and elevated groundwater temperature. In Phase 3 of the Play Fairway project, drilling just outside the Pālāwai Basin confirmed increasing water temperature to a depth >1 km (Lautze and Thomas, 2021). The prospectivity of geothermal resources outside of the LERZ on the Island of Hawai'i remain largely unknown. Current trends in geothermal energy focus on the use of Machine/Deep Learning (ML) for industry problem-solution and decision-making (Smith et al., 2022).

In general, ML applications fall into three classes: data-driven ML, physics-informed ML, and multimodal ML. According to Qin et al. (2022), data-driven ML models potentially suffer from three limitations: 1) extensive data requirements, 2) lack of physical plausibility and interpretability, and 3) poor generalizability beyond training data. Physics-informed ML can overcome these shortcomings by either 1) embedding geophysical or mass and/or energy balance equations into the ML algorithm, or 2) informing the ML algorithm using properties derived following the inversion of (multi)physics-based equations. In the (multi)physics-informed ML approach, the incorporation of physics into the ML algorithm typically relies on using simplified set of geophysical or mass and/or energy balance equations (He et al., 2020) along with the addition of a physics-guided loss function to regularize training of the associated parameter types. In an alternative approach, a complete set of explicitly coupled water-heat-solute transport equations are used together with supervised ML algorithms that stochastically sample different types of field measurements to reduce the computational burden and improve the parameter estimates (Friedel, 2023).

In this study, we present a multimodal machine learning (MML) workflow to assimilate, discover, and predict the distribution of numeric and categorical features (field observations) along a groundwater-geothermal continuum. Success of the MML workflow relies on a transductive learning algorithm that maximizes mutual information (measure of entropy describing mutual dependence among random variables) when projecting disparate and incomplete field observations onto a single embedding space (hypersurface). Multimodalities can include any combination of *measured* (point field) and *derived* (multiphysics-based numerical model inversions, data-driven machine learning, and multiphysics-informed machine learning) *features.* The aim of this study is to demonstrate efficacy in using the MML workflow for identifying hidden groundwater and geothermal resources and attendant geologic features in Hawai'i. We hypothesize that the MML can assimilate and predict heterogeneous and incomplete features (modalities) for locating hidden resources by exploiting mutual information in multimodal field observations. To test this hypothesis, the proposed MML workflow is applied to assimilate subsets of Hawai'i Play Fairway modalities and predict subsurface geophysical, geologic, and hydrogeologic features across the Islands of Lāna'i and Hawai'i. The successful application of the MML workflow will achieve the following objectives: 1) develop open community datasets, 2) identify data acquisition targets with high value for future work, 3) identify new signatures to detect hidden groundwater and geothermal resources, and 4) foster new capabilities for characterizing subsurface temperature and permeability. This study extends the work of Friedel (2016) who applied a MML workflow to construct a continuous set of hydrostratigraphic units (HSU) for characterizing shallow (<200m) groundwater systems, and Vesselinov et al. (2020) who used an unsupervised ML methodology to discover hidden signals in field data and extract their dominant attributes in a shallow (<200m) hydrothermal system.

## 2. METHODS

The MML workflow used to determine 3d Geothermal Stratigraphic Units (GSUs) involves three primary steps: (1) feature selection, (2) feature prediction, and (3) feature clustering.

### 2.1 Feature Selection

The identification of suitable features (observation types) for building the MML model is undertaken using the nonlinear wrapper approach called learn heuristics with feature constraints. The wrapper model requires one predetermined learning algorithm in feature selection and uses its performance to evaluate and determine which features are selected (Yu and Liu, 2003). In this regard, the learn heuristics reflect the ML algorithm that is introduced into the Metaheuristics algorithm (Buscema et al., 2013). The ML algorithm is generalized for which multiple learning algorithms are evaluated, for example, Backpropagation, K-Nearest Neighbor, or Naïve Bayes, inside a genetic algorithm (GA) (metaheuristic). Constraints on the learn heuristic approach typically involve features such as one or more response variables. Given that the GA converges to a global minimum surface (not a single vector), the process is randomly reinitiated and run until the subset of optimal features can be identified based on their mode of values.

### 2.2 Feature Prediction

A Modified Self-Organizing Map (MSOM) procedure is used to predict feature vectors at unsampled locations. The prediction method is sufficiently robust to cope with feature vagaries due to sample size and extreme data insufficiency, even when >80% of the data are missing (Friedel and Daughney, 2016). The transductive MSOM procedure involves the sequential application of competitive learning (self-organizing map) and estimation (minimization of a two-component objective function by competitive learning). The aim is to maximize mutual information between the input and latent code vectors. The architecture of MSOM involves an input layer (signals from the environment) and an output layer (competitive feedback to the environment). The input layer comprises a set of nodes (neurons) that are connected one to another through a rectangular topology (rows by columns). The connections between inputs (data vectors) and nodes have weights, so a set of weights corresponds to each node.

In implementing the MSOM procedure, the competitive learning process iteratively modifies weights during the training phase so that the self-organized output pattern becomes consistent, meaning that the input pattern produces the same output pattern. In doing so, the MSOM iteratively maps each data sample as a vector (each variable is characterized as a cloud of data vectors) across a hypersurface on which data vectors closer to each other are more related (self-similar) than data vectors farther away. The learning algorithm may be summarized as follows (Kohonen, 2001):

1. Generate initial values including weights, radius, and learning parameter values
2. Select input vector from data set
3. Identify winning node, which is the closest node to the input vector using the Euclidean distance metric
4. Identify the neighborhood with the given radius using a Gaussian function
5. Update the weights for the winning node and all the nodes in the same neighborhood
6. Repeat steps 2 to 6 until the weight vectors reach a converged state
7. The estimation of missing values (sometimes referred to as imputation) is done simultaneously for all variables across the hypersurface (Kalteh et al., 2008).

Feature prediction is undertaken for each new instance, x, by searching through the MSOM best-matching unit vectors for the K most similar instances (Wang, 2003; Riese, 2019). To determine which of the K instances in the training dataset are most like the new instance, the Euclidean distance metric is used while iteratively minimizing the sum of quantization and topographical error vectors as part of the competitive learning process (Kalteh et al. 2008).

Training and testing of the MSOM model are carried out using a split-sample validation approach. In this approach, the data records are randomly shuffled and split with 80% set aside for training and 20% set aside for testing. In total, the original data set is split from the original data set using this procedure N times. Each split characterizes a fold used in computing the N-fold cross-validation statistics to assess the generalizability of the unsupervised ML algorithm when presented independent field data. The metrics used to evaluate model performance include R-squared (a statistical measure representing the proportion of variance for the observed variable that is *explained* by the estimated variable) and mean squared error (predictive success) for continuous features; and accuracy (predictive success) and Cohen's Kappa (comparison between the observed accuracy and the agreement expected due to chance) for categorical (present or absent) features (Cohen, 1960). Cohen suggested the Kappa result for categorical features be interpreted as follows: values ≤ 0 as indicating no agreement and 0.01–0.20 as none to slight, 0.21–0.40 as fair, 0.41– 0.60 as moderate, 0.61–0.80 as substantial, and 0.81–1.00 as almost perfect agreement.

### 2.3 Feature Clustering

Grouping statistically meaningful features across the self-organized hypersurface is undertaken by k-means clustering (Vesanto and Alhoniemi 2000) of information across the learning network. The best partitioning for each number of clusters is determined based on the Euclidian distance criterion and interesting merges defined using the Davies-Bouldin index (Davies and Bouldin 1979). In this study, the optimal number of clusters is determined by repeating the k-means process to avoid convergence at local minima. The Davies-Bouldin validity index is selected as a matter of convenience, but other validity measures could be used.

## 3. CASE STUDIES

The MML study focuses on the Islands of Lānaʻi (Friedel et al., 2022) and Hawai'i briefly described in the following sections.

### 3.1 Island of Lānaʻi

#### 3.1.1 Study Region

The Island of Lānaʻi and adjacent Pacific Ocean is located about 96 km southeast of Honolulu, Oahu (Figure 1). Lānaʻi is a cashew-shaped island with a maximum width of 29 km in the longest direction. The highest point on Lānaʻi is 1026 m with a land area of 364 $km^2$. The geology of Lānaʻi is described as an eroded extinct basaltic volcano that developed during one period of activity (Stearns, 1940). Structurally, the island has three primary rift zones (Northwest Rift zone, Southwest Rift zone, and South Rift zone) and caldera located in the Pālāwai Basin (Figure 2). The summit plateau resulted from collapse along the northwest rift zone. Lānaʻi island has fault breccias and dike complexes that lie in the rift zones radiating from the Pālāwai Basin. Relatively low permeability dikes often enclose lavas of relatively high permeability forming local compartmentalized reservoirs (Stearns,1940). Basaltic rocks exposed along the west coast are thinly bedded and considered the most permeable rocks due to cavities and fractures within and between lava flows and thought to permit the landward intrusion of sea water. Groundwater temperatures measured in a well on the rim of the Pālāwai Basin during Play Fairway Phase 3 vary from about 21 C at the land surface to about 65 C at a depth of 1 km. Little is known about the subsurface groundwater-geothermal system outside these locations, although temperatures at the Moho (about 13 km depth) are likely in the range of 444 C to 892 C (Schutt et al., 2018). Therefore, understanding the occurrence of groundwater and geothermal resources requires knowledge of the spatial distribution and characteristics of these basalt-dike systems.

#### 3.1.1 Field Data

Field data used in the Lānaʻi case study reflects a subset of those data assembled and/or collected during the Hawaiʻi Play Fairway project, funded by the DOE's GTO. This data set includes geologic, groundwater, and geophysical observations relevant to subsurface heat, fluid, and permeability for collected at the island of Lānaʻi (Lautze et al., 2020). For example, geologic data includes information on basalt and dike rocks identified at the surface and in drill cores. Groundwater data collected in boreholes include physical properties (water level and specific capacity), aqueous chemistry (major ions, nutrients, and metals), environmental tracers (isotopes), and aqueous parameters (dissolved oxygen, specific conductivity, and temperature). These datasets reflect field sampling from the surface to a maximum borehole depth of about 1 km (Figure 2). To connect this sample region to greater depths, un-collocated surface gravity and magnetotelluric measurements were collected and deterministically inverted to estimate 3d distribution of density and electrical resistivity values across Lānaʻi to depths of more than 14 km (Lautze et al., 2020).

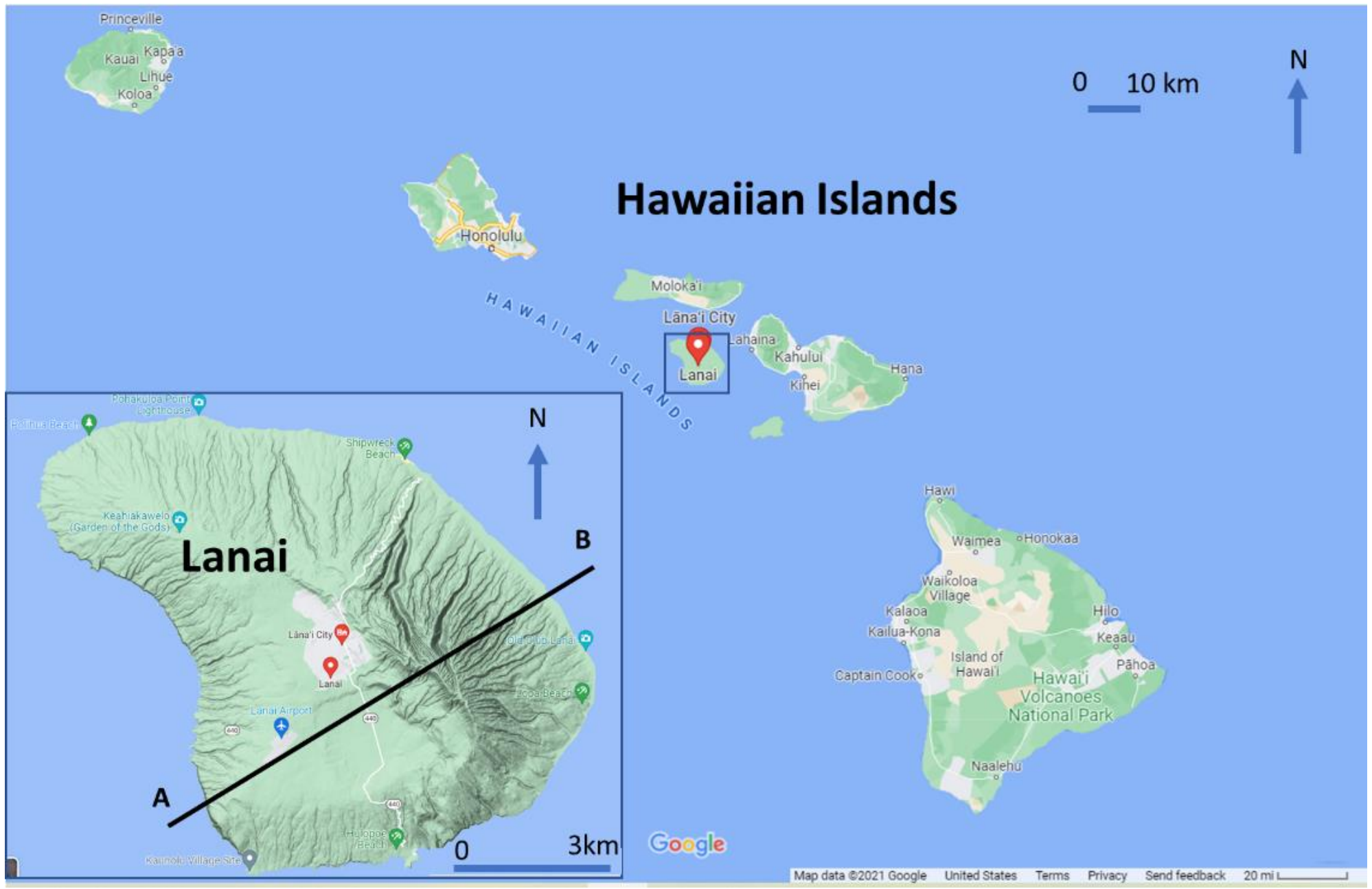

**Figure1. Lānaʻi study map showing proximity of Lānaʻi to the Hawai'ian Islands and Lānaʻi cross-section (inset) for which MML results are extracted (from the surface to 15 km) for evaluation and construction of the conceptual model.**

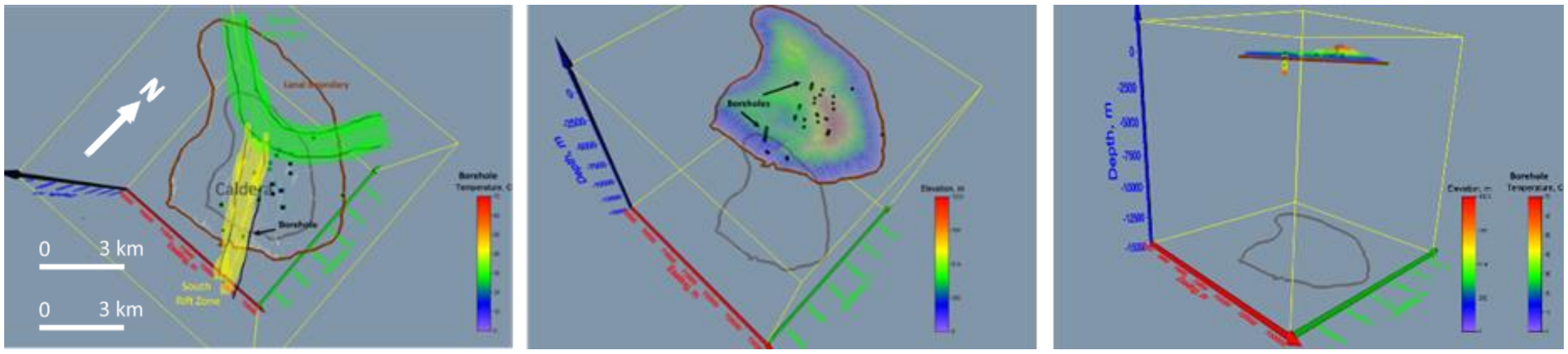

**Figure 2. Study map and model framework showing Lānaʻi coastal boundaries at sea level (brown) at 15km depth (gray); left panel includes primary rift zones: north (green) and south (yellow); caldera, borehole/well locations (shown as dots with color indicating temperature); middle and right panels show location of borehole/well locations with respect to elevation.**

### 3.2 Island of Hawai'i

3.2.1 Study Region

The Island of Hawai'i, also referred to as the "Big Island," is largest of the Hawaiian Islands (Figure 3). At its greatest dimension, the island is 150 km (93mi) across with a total land area of 10,432 km$^2$ (4,028 mi$^2$). The Big Island has five subaerial (above sea level) volcanoes: Hualalai, Kilauea, Kohala, Mauna Loa, and Mauna Kea. Of these volcanoes, Kohala is extinct and oldest with an estimated age of 1 million years. In contrast, Hualalai and Mauna Kea are dormant, and Mauna Loa and Kilauea are active. The most recent eruption of Mauna Loa occurred in 1984 whereas Kilauea has been erupting regularly since 1983. The highest point on the Big Island is the dormant volcano Mauna Kea 4,205 m (13,796 ft).

The geology of the Big Island reflects four stages of volcanism (Stearns, 1946) as identified from surface and subsurface rock samples. Specially, during the initial submarine stage (stage 1) small vents or fissures open in the ocean floor and erupt pillow lavas that have higher relative abundances of sodium and potassium than the tholeiitic basalts of the shield-building stage. Continued volcanism produces a shield volcano. As the top of the volcano approaches sea level (stage 2), the water pressure at the volcano summit drops allowing water in contact with lava to turn to steam, resulting in explosions that form tephra and hyaloclastite rocks. In the subaerial shield-building stage aa and pahoehoe erupt from the summit area and rift zones resulting in growth of the volcano. In the subaerial landslide stage (stage 3), large volumes of unstable volcanic mass (slump) form toward the coastal edges of the volcano being translated seaward under their own weight eventually failing as a landslide and/or debris flow into the adjacent ocean (Moore et al., 1994). Kilauea appears to be in this stage with the largest of Hawai'i's historic earthquakes resulting from movement of the Hilina slump in the southern part of the Big Island. In the capping stage (stage 4). alkalic basalt fills the summit caldera and produces a steep-sided cap at the summit of the volcano. Cinder cones develop from particles and blobs of congealed lava ejected from a single vent. Mauna Kea is an example of a volcano with multiple steep sided (25-33 degrees) cinder cones.

Structurally, Hawai'i island has four primary rift zones: Southwest Rift Zone and Northeast Rift zone (referred to herein as the Mauna Loa Rift), Southwest Rift zone and East Rift zone (referred to herein as the Kilauea Rift). Both the SW Rift zone and East Rift zone bound the north side of the Hilina Fault system. The Hilina Fault system is considered the northern boundary of the Hilina Slump, a 19,800 $km^3$ (4,760 $mi^3$) section of the south slope of the Kilauea volcano which is moving away from the island (Delinger et al., 2014). The Hilina Slump, on the southern flank of the Kīlauea Volcano on the southeast side of the island of Hawai'i, extends from the Hilina fault zone approximately south of the East Rift Zone (ERZ) to the edge of deep water (Delinger et al., 2014).

The rocks of Hawai'i island are considered highly permeable resulting in rapid rainfall infiltration into the subsurface (Stearns, 1946). Perennial streams are present only on the western slopes of Kohala Mountain and Mauna Kea. Most of the water infiltrates rapidly to the basal water table, where the freshwater floats on salt water according to the Ghyben-Herzberg principle. Basal water discharges in springs at or near sea level all along the coast. Only a very small proportion of this groundwater is recovered in wells. Along the western coasts the basal water is of good quality and large supplies await development. Along the eastern coasts most of the basal water is brackish.

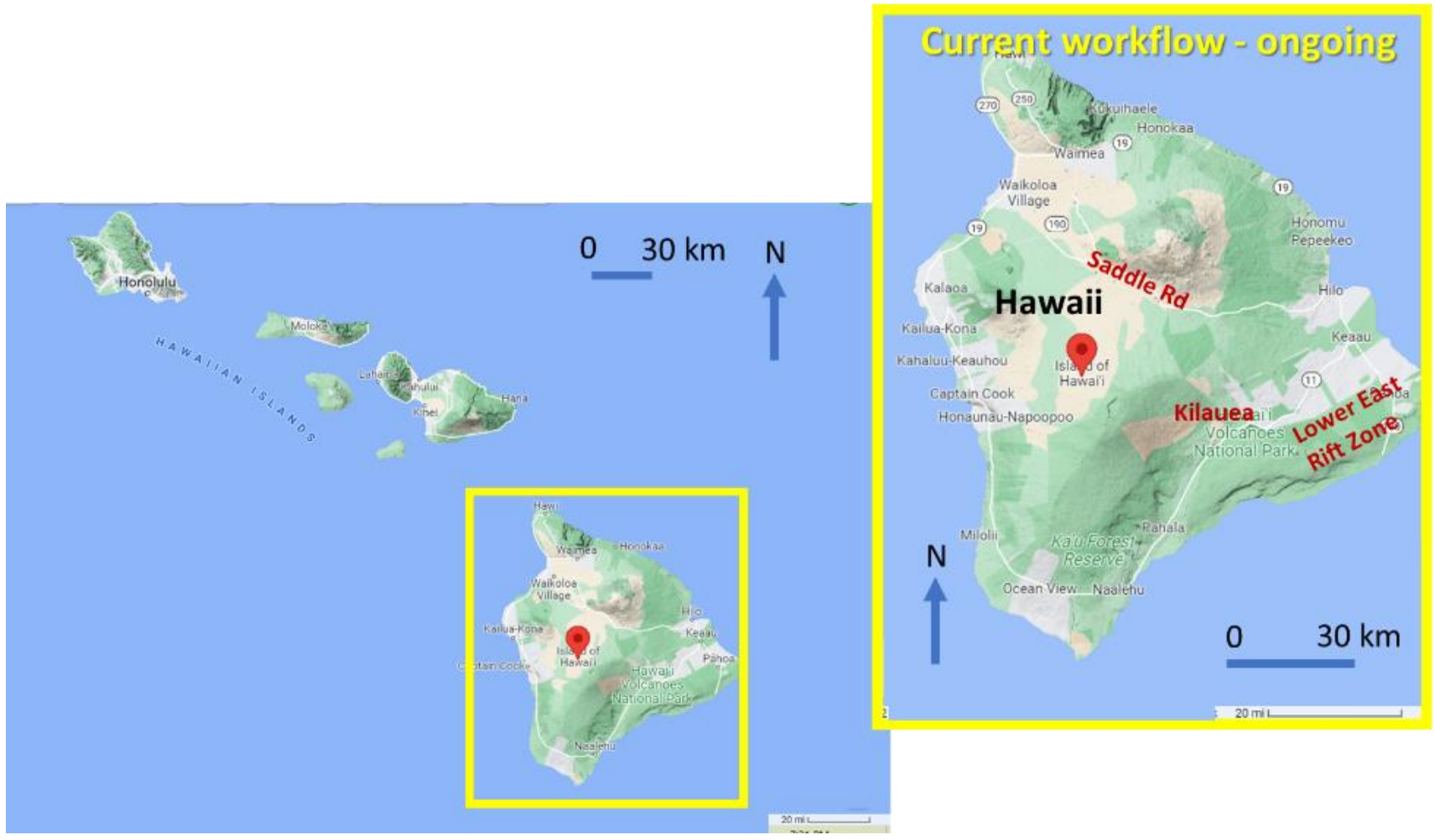

**Figure 3. Hawai'i study map showing (left) proximity of the Island of Hawai'i to the Hawai'ian Islands, and (right) the Saddle Rd, Kilauea, and Lower East Rift Zone study areas for which MML is used to assimilate and predict features locally (individual areas), regionally (three areas) and island-wide (three areas and random locations of pseudo-boreholes from surface to 40 km depth).**

3.2.2 Field Data

Field data used in the Hawai'i case study reflects a subset of those data assembled and/or measured during The Hawai'i Play Fairway project, funded by the DOE's GTO along with publicly available seismic data (Incorporated Research Institutions for Seimology - IRIS, 2022), and various published and contractor reports (Jerran et al., 2019, State of Hawai'i, 1990). This data set includes observations from four feature categories: location (e.g., Easting, Northing, and Elevation); state variable (e.g. temperature) (Putrika, 1997; Lautze et al., 2020)); Geology: Hilina, Kahuku, Kaoiki, Kealakekua, Koae, and Kohala fault systems; Kilauea and Mauna Loa rifts (Lipman et al, 1990; Wolfe and Morris; 1996; Trusdell et al., 2006); basalt, oceanic crust, underplating, mantle (Leahy et al., 2010); and physical properties (Lautze et al., 2010), such as, density, resistivity, p-wave velocity, s-wave velocity, earthquake magnitude.

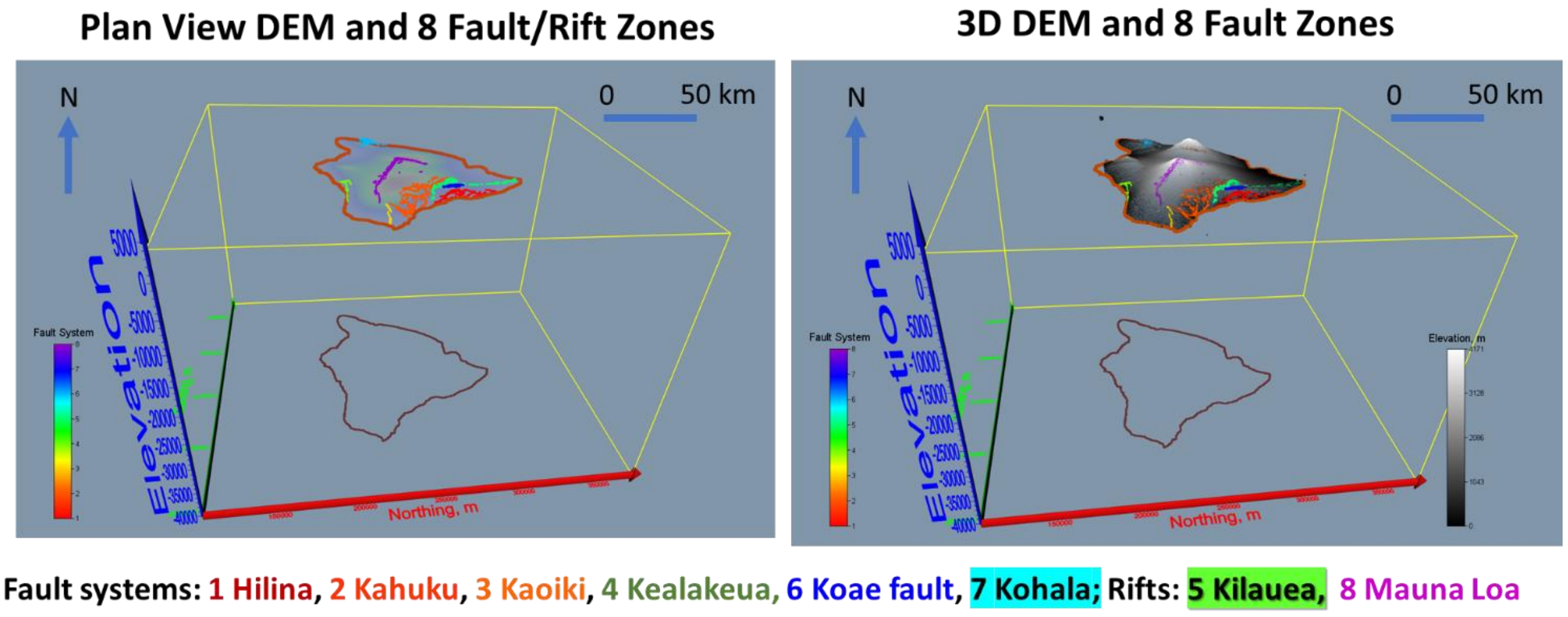

**Figure 4. Study map and model framework showing Hawai'i coastline at sea level (orange) and its projection at 40 km depth (brown); left panel includes primary fault/rift systems 1 to 8 and layered on plan view dem (e.g., Fault systems: 1 Hilina (red), 2 Kahuku (orange-red), 3 Kaoiki (orange), 4 Kealakeua (dark green), 6 Koae fault (royal blue), 7 Kohala (light blue); and Rifts: 5 Kilauea (light green) and 8 Mauna Loa (purple); the right panel includes primary faults zones draped onto the dem from 0 (sea level) to 4171 m.**

The Hawai'i temperature data reflect measurements collected from shallow (generally 25-100 m) groundwater wells (N=32) along the island coast and geothermal boreholes in the Saddle Rd (N=2). Kilauea (N=1) , and LERZ (N=9) areas (Lautze et al., 2020). The groundwater well data reflect field sampling from the surface to a maximum borehole depth of about 1 km (Figure 2), whereas the geothermal wells reflect measurement depths as great as 3 km. To connect the groundwater and geothermal sample areas to greater depths, uncollated surface gravity, magnetotelluric and seismic measurements were collected and deterministically inverted to estimate 3d distribution of density, electrical resistivity, and velocity values across study areas to depths up to 40 km. A spatial overview of sample locations for these study areas is presented in Figure 5a.

At the Saddle Rd study area (Figure 5b), surface audiomagnetotelluric and magnetotelluric measurements were recorded and deterministically inverted to provide 1d density and 1d resistivity (Constable et al., 1987) profiles at 31 stations (12PT, 13PT, 15PT, 16PT, 1PT1, 2PT1,3PT1, 4PT1, 5PT1, 6PT1, 7PT1, 8PT1, 9PT1, HH01, HH02, HH03, KK01, KK02, KM01, MK1, ML01, ML02, MM13, OR01, PK01, PK02, Pk03, PK04, PK05, REF1, and WF01) from the surface to a depth of 3 km, and the same measurements were also inverted using stochastic inversion scheme developed as part of this project to provide 1d resistivity profiles spanning deciles from the surface to a depth of 3 km. Another deterministic inversion was undertaken using the same magnetotelluric measurements to provide 1d resistivity profiles to a depth of 40 km ensuring the project had coverage spanning the Moho (Leahy et al., 2010). A separate deterministic gravity inversion (UBC REF) provided density measurements along surface stations to a depth of 3 km. In addition to surface measurements, other temperature, velocity, and resistivity measurements recorded to a depth of 1.5 km from the PTA2 and KMA1 boreholes also were used in this study.

At the Kilauea study area (Figure 5c), surface magnetotelluric measurements were recorded and deterministically inverted (Constable et al., 1987) to provide 1d resistivity profiles at 16 stations (g1sta03rrhkml, g1ta04rrhkml, g1sta05rrhkml, g1sta19rrhkml, g2sta13rrhkml, g2sta1rrhkml, g3sta12rrhkml, , g4sta01rrpark, g4sta06rrpark, g4sta08rrpark, g5sta07rrpark, g5sta11rrpark, g6sta09rrpark, g6sta101rrpark, g6sta17rrpark, and g6sta18rrpark) from the surface to a depth of 40 km. Other seismic measurements located within 150 m of the magnetotelluric stations were recorded and inverted as part of this project to obtain discrete p-wave and s-wave velocities to depths exceeding 40 km. A single borehole sponsored by National Science Foundation (Keller-NSF) provided temperature measurements to a depth of 1.5 km.

At the Lower East Rift Zone (LERZ) study area (Figure 5c), surface magnetotelluric measurements were recorded an deterministically inverted (Constable et al., 1987) to provide 1d resistivity profiles at 9 stations (km01, km02, ps02, ps08, ps09, ps11, ps12, and ps19) from the surface to a depth of 40 km. Seismic measurements located within 150 m of the magnetotelluric stations were inverted as part of the project to obtain p-wave and s-wave velocities to depths exceeding 40 km. Measurements from several geothermal boreholes (Ashida 1, Geothermal 1, Geothermal 2, HGP-A, KS-1, KS-2, Lanipuna-1, SOH1, SOH2, Geothermal 4, SOH4, Keauohana 1, and Malam Ki) provided temperatures from the to a depth of about 1.5 km (Jerran et al., 2019, Stolper et al., 2009, Sorey and Colvard, 1994; Puna Geothermal Venture, 1991, State of Hawai'i, 1990, Campbell and Gardner, 1981; Kihara et al., 1977; Kingston Reynolds Thom & Allardice Ltd., 1976). In addition, measurements from a single geothermal well north of the LERZ near Hilo (Stolper et al., 2009) provided temperatures at the coastal boundary.

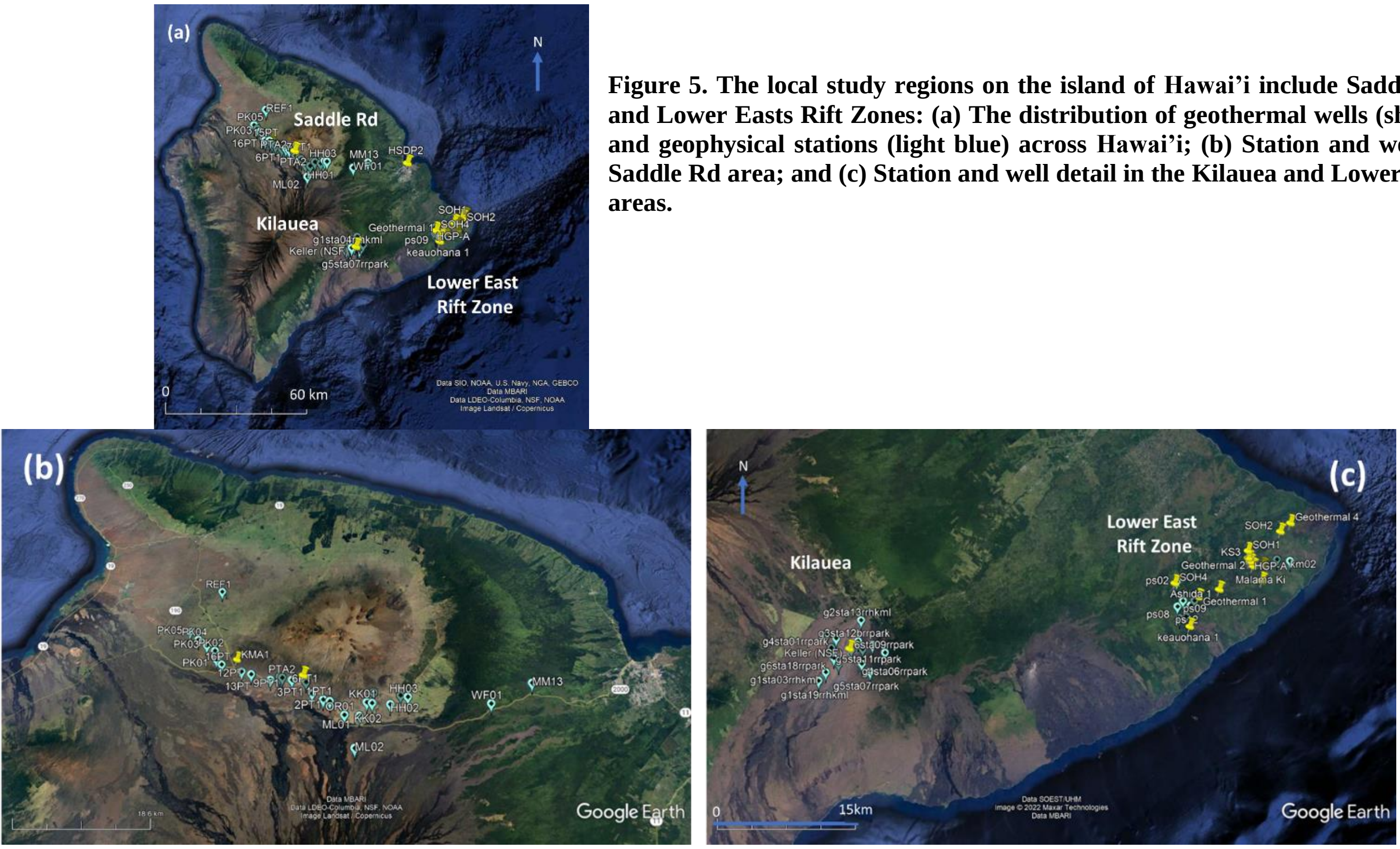

**Figure 5. The local study regions on the island of Hawai'i include Saddle Rd, Kilauea, and Lower Easts Rift Zones: (a) The distribution of geothermal wells (shown in yellow) and geophysical stations (light blue) across Hawai'i; (b) Station and well detail in the Saddle Rd area; and (c) Station and well detail in the Kilauea and Lower East Rift Zone areas.**

## 4. PRELIMINARY RESULTS

### 4.1 Island of Lānaʻi

<u>4.1.1 Feature Selection</u>

Of the original number of predictor variables (N = 54), the learn-heuristic based feature-selection process identified an optimal global set of variables (N = 27) suitable for predicting features beneath Lānaʻi (Table 1). These variables are used in building of the unsupervised learning-based models. Important predictors that appear common to all three state variables include Water: Ocean, Geology: basalt and dike, Physical properties: density, resistivity, specific capacity; Aqueous properties: specific conductivity, oxygen reduction potential, and pH; Aqueous chemistry: HCO3, NO3, PO3, SO4, r, F, F, K, Na, Si, Sr, and Ca/Mg ratio.

**Table 1. Summary of features determined to be informative using the learn heuristics approach constrained to the state variables of head (m), temperature (C), and chloride concentration (mg/l). Physical properties: resistivity (ohm-m), specific capacity (m3/d/m); Aqueous properties: specific conductance, uM/s, oxygen reduction potential (mv); Aqueous chemistry: Br = bromide, (mg/l) F = Fluoride (mg/l), Fe = iron (mg/l), HCO3 = bicarbonate (mg/l), K = potassium (mg/l), Na = sodium (mg/l), NO3 = nitrate (mg/l), PO4 = phosphate (mg/l), Si = silica (mg/l), SO4 = sulfate (mg/l), Sr = strontium (mg/l), Ca/Mg = calcium -magnesium ratio.**

Constraints on learn heuristics

Informative features

| Category | Features | Type: Discrete | Type: Continuous | Observation: Measured | Observation: Derived | Support: Point |
|---|---|---|---|---|---|---|
| State variables | Head, m | | X | X | | X |
| | Temperature, C | | X | X | | X |
| | Chloride concentration, mg/l | | X | X | | X |
| Water | Ocean | X | | X | | X |
| Geology | Basalt | X | | X | | X |
| | Dike | X | | X | | X |
| Physical properties | Density, kg/m3 | | X | | X | |
| | Resistivity, ohm-m | | X | | X | |
| | Specific capacity, m3/d/m | | X | | X | |
| Aqueous properties | Specific conductivity | | X | X | | X |
| | Oxygen reduction potential, mv | | X | X | | X |
| | pH | | X | X | | X |
| Aqueous chemistry | HCO3, mg/l | | X | X | | X |
| | NO3, mg/l | | X | X | | X |
| | PO4, mg/l | | X | X | | X |
| | SO4, mg/l | | X | X | | X |
| | Br, mg/l | | X | X | | X |
| | F, mg/l | | X | X | | X |
| | Fe, mg/l | | X | X | | X |
| | K, mg/l | | X | X | | X |
| | Na, mg/l | | X | X | | X |
| | Si, mg/l | | X | X | | X |
| | Sr, mg/l | | X | X | | X |
| | Ca/Mg | | X | X | | X |

4.1.2 Feature Prediction

Prior to training of the MSOM, features (field variables) were normalized by their data variance and randomly assigned (presenting the input vectors to the map sequentially using a randomly sorted database) as an initial set of map weight vectors. Application of the MSOM network to training data is done using a single fixed number of neurons and topological relations. The selected neural map shape (148 rows by 140 columns) is a toroid (wraps from top to bottom and side to side) with hexagonal neurons. Training of the map was conducted using both rough and fine phases. The rough training phase involved 20 iterations using a Gaussian neighborhood with an initial and final radius of 204 units and 51 units; and the fine training involved 400 iterations using a Gaussian neighborhood with an initial and final radius of 51 units and 1 unit. The initial and final learning rates of 0.5 and 0.05 decayed linearly down to $10^{-5}$, and the Gaussian neighborhood function decreased exponentially from a decay rate of $10^{-1}$ iteration to $10^{-3}$, providing reasonable convergence evidenced by similarity in their low quantization ($q_e$=0.073) and topographic ($t_e$=0.076) errors.

Testing and validation of the trained MSOM model are reflective of the five split sets previously discussed. The cross-validation statistics further reveal that the MSOM model has a moderate ability (kappa values > 0.53) to predict dikes and basalt with strong prediction accuracies (accuracy values > 0.99). For the sake of brevity, summary statistics and scatterplots are presented for independent observations and predictions of selected features that include temperature, chloride concentration, resistivity, and density (Figure 6). Note that the metrics for temperature predictions are reflective of those measured in the upper 1 km, because the magnitude of increasing temperatures at greater depths is unknown. In general, the cross-validation statistics (table 2) reveal similar observed to predicted values (minimum, average median and maximum) with reasonable prediction accuracies (standard deviation / square root of count): resistivity = 0.32 ohm-m, density = 1.2 kg/m3, temperature = 0.002 C, chloride concentration = 1.16 mg/. The associated scatter plots (Figure 3) of observed versus predicted values for these features have reasonable R-squared values (resistivity, ohm-m = 89.9%; density, kg/m3 = 99.9%, chloride, mg/l = 68.8%; temperature, C = 92.8%) albeit with some bias at smallest values. The bias at small values for density, conductivity, and chloride concentration is attributed to challenges in predicting these features at the ocean-basalt interface given the large voxels used during deterministic inversions and moderate ability to predict the presence or absence of basalt. Future studies may benefit by introducing higher frequency audiomagnetotelluric measurements to improve near surface estimates of resistivity at the coastal boundary and shallow depths. Likewise, the prediction of small density values may be reflective of the inability of MSOM to resolve density at the ocean-coastal boundary due to grid resolution.

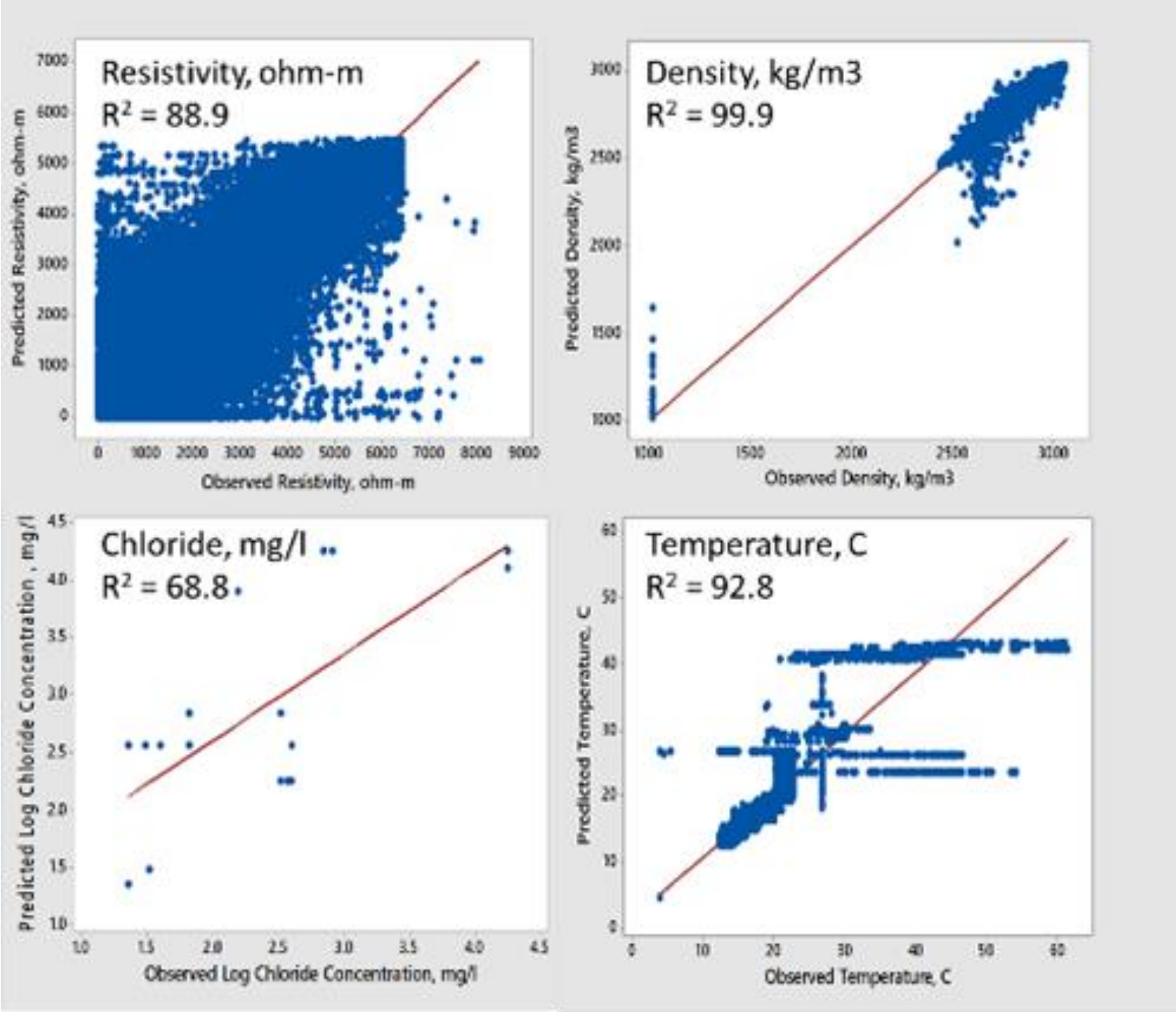

**Figure. 6. Scatterplots showing observed versus predicted values for selected features with fitted line and R-squared values: resistivity, density, chloride concentration and temperature.**

**Table 2. Summary statistics for observed and predicted features: temperature (C), chloride concentration (mg/l), resistivity (ohm-m), and density (kg/m3). Note that statistics for the temperature are confined to those measured in the upper 1 km given that temperatures at the Moho are assumed to be about 758C (Schutt et al., 2018).**

Temperature, C

| Statistic | Observed, C | Predicted, C | Error, C |
|---|---|---|---|
| Minimum | 4.00 | 4.00 | 0.00 |
| Average | 22.5 | 23.4 | 0.34 |
| Median | 22.4 | 23.4 | 0.03 |
| Maximum | 61.4 | 61.3 | 30.4 |
| Standard deviation | 4.69 | 8.19 | 1.22 |
| Count | 213824 | 1044532 | 213824 |

Chloride, mg/l

| Statistic | Observed, mg/l | Predicted, mg/l | Error, mg/l |
|---|---|---|---|
| Minimum | 23.0 | 23.0 | 0.0 |
| Average | 17999 | 12956 | 1.3 |
| Median | 18000 | 18000 | 0.0 |
| Maximum | 18000 | 18002 | 17670 |
| Standard deviation | 157.7 | 7974 | 533.3 |
| Count | 212919 | 1044530 | 212919 |

Resistivity, ohm-m

| Statistic | Observed, ohm-m | Predicted, ohm-m | Error, ohm-m |
|---|---|---|---|
| Minimum | 0.10 | 0.14 | -86.5 |
| Average | 516 | 385 | 130 |
| Median | 100 | 35.1 | 25.4 |
| Maximum | 8066 | 5470 | 7181 |
| Standard deviation | 940 | 786 | 288 |
| Count | 799480 | 1044530 | 799480 |

Density, kg/m3

| Statistic | Observed, kg/m3 | Predicted, kg/m3 | Error, kg/m3 |
|---|---|---|---|
| Minimum | 1020 | 1020 | 0.00 |
| Average | 1129 | 2016 | 21.9 |
| Median | 1020 | 2617 | 0.00 |
| Maximum | 3055 | 3037 | 2029 |
| Standard deviation | 420.6 | 862.4 | 182.6 |
| Count | 227176 | 1044530 | 227176 |

A natural outcome of using the trained MSOM model is the simultaneous prediction all model features. Given the large number of available features, a reduced number of features is selected for presentation that include state variables (temperature, chloride concentration), physical properties (specific capacity – a surrogate for permeability), geology (basalt and dike), and Geothermal Stratigraphic Units (GSU). Each of these features represent predicted values extracted from the 3d data cube along the cross-section A-B. The cross-section A-B striking SW-NE along a profile that begins at the Pacific Ocean on the west, crosses the caldera dike system, the point of highest elevation, and ends at the Pacific Ocean on the east. Collectively, figures 7-11 reveal a heterogeneous groundwater – geothermal system from which information can be useful in developing a conceptual model. Note that the model domain boundary is indicated as having poor resolution given the lack of available field data in these regions. In the balance of the model domain, testing reveals that the prediction process preserves reasonable population statistics for the geologic, hydrogeologic and geophysical features, but the individual feature predictions reflect a random process (statistics change for different split sets). For this study, the cross-validation statistics demonstrate the ability of the MSOM to generalize when estimating continuous features from sparse field data characterizing the different support volumes.

This section presents a subset of features predicted while using the trained MSOM and 5 sets of independent observations. For example, the 3d density and resistivity predictions are presented in Figure 7. Some general spatial observations are that both density and resistivity are heterogeneous through the region. The interpreted density anomalies include minimum values associated with the ocean and greatest density inland at the caldera and to depth of about 15 km. The predicted resistivity also appears heterogenous through the region with features at sea level displaying more resistive character at and below the region of greatest elevation and conductive features similar in magnitude to the ocean but decreasing landward in areas located at the northern most point and southwestern point (toward point A on the cross section). Another slightly deeper region of high conductivity (low resistivity) appears at the eastern boundary of the island (toward point B on the cross section). These areas suggest regions landward intrusion of sea water with decreasing gradients over about 5 km. In the area of maximum elevation, the resistivities are greatest possibly due in part to groundwater recharge of freshwater. Below about 1 km, the anomalies appear to be vertically oriented as rectangular regions of alternating medium to high resistivity values.

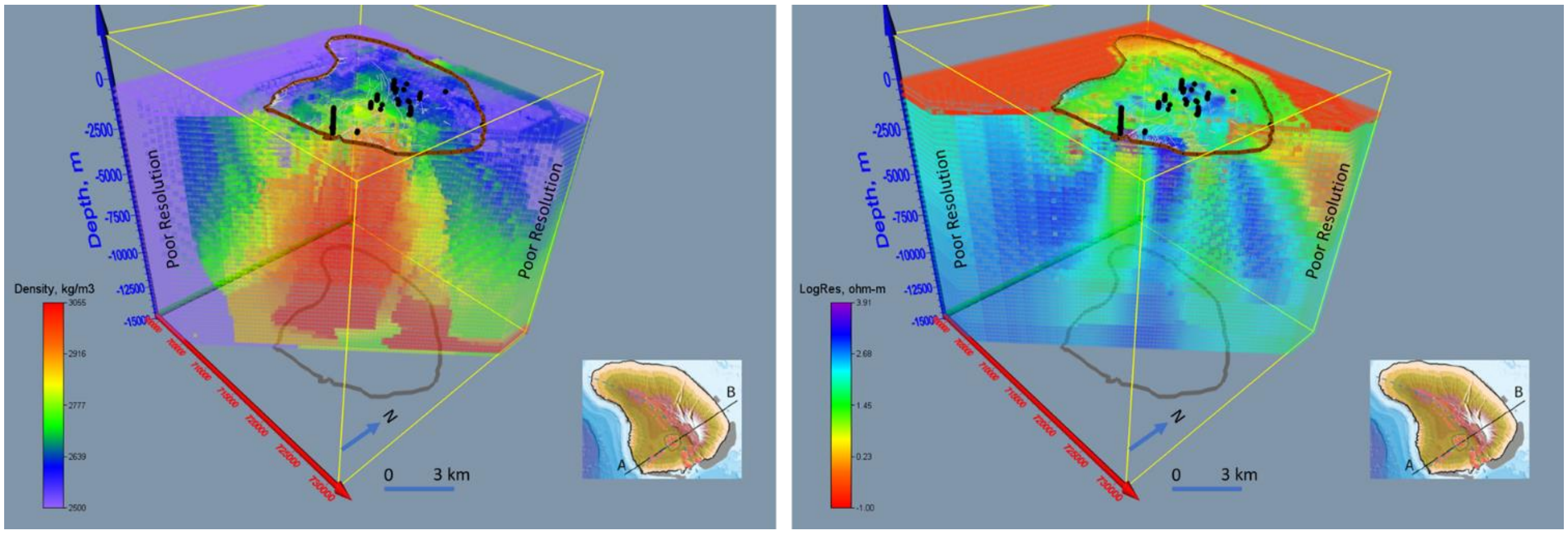

**Figure. 7. Inverted 3d geophysical property distributions sliced horizontally at sea level and vertically along A-B: (left) density, kg/m3 (density > 2500 kg/m3), and (right) log resistivity, ohm-m. Regions of poor resolution reflect limited field data available to inform the model.**

The predicted spatial occurrence of subaerial basalt, referred to as basalt, (left) and submarine basalt and dike, referred to as dike, (right) are shown in Figure 8. The general character of the predicted basalt is the prominence from surface to depths as great as about 10 km. One characteristic that the basalt appears to be draped over the region indicated as not basalt. The not basalt region appears as dike material with several interesting features. In fact, the dike material in the region of not basalt is interpreted as a domed shape pluton with roots below the Moho located at about 12.5 km depth. The change in character from the pluton at depth is interpreted as the possible location of the Moho. Other vertical features are interpreted as rising dike swarms to the west side of the pluton (between 5 and 8 km depth) and at the top of the pluton below the caldera region (between sea level to about 1 km depth). The later dike swarms correspond to the region denoted as a dike zone by Malahoff and Woollard (1966) based on airborne total field magnetic anomalies. Lastly, the western dike swarm rose vertically to a depth of about 5 km below the surface where they terminate forming what appears to be a sill.

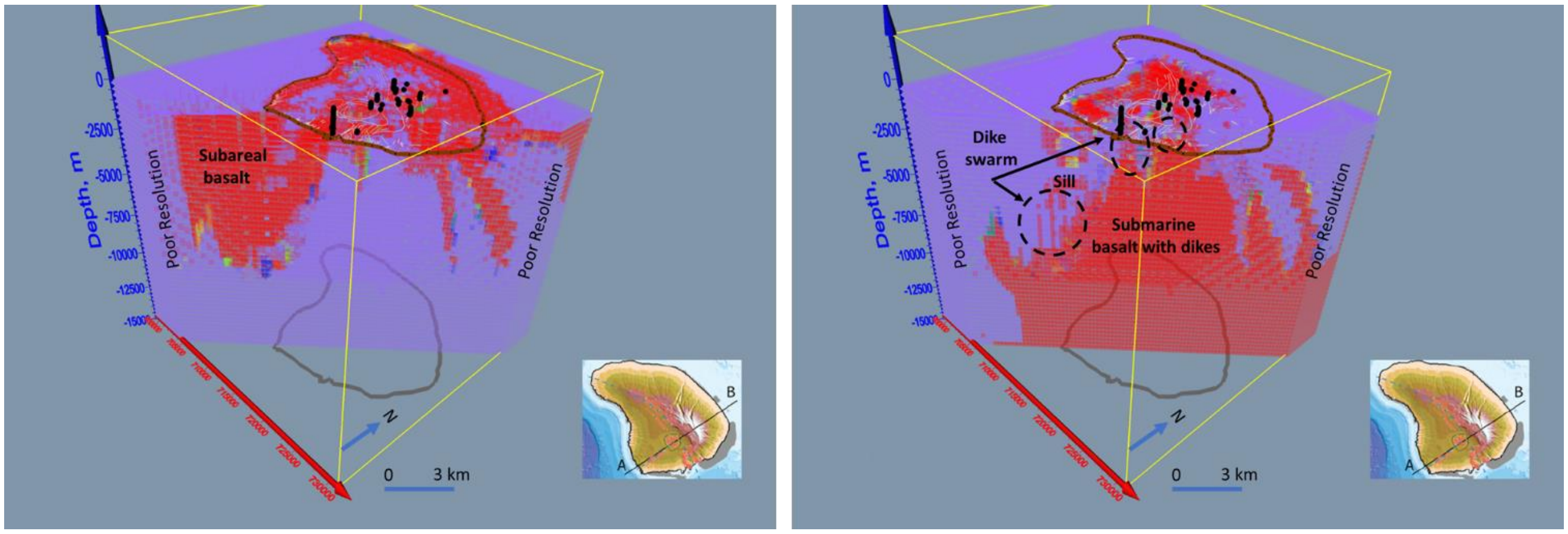

**Figure 8. Predicted 3d geology distribution (95% likelihood to be present or absent) sliced horizontally at sea level and vertically along A-B: subaerial basalt, referred to as basalt (left), and submarine basalt and dikes, referred to as dike (right). Dike image includes likely dike swarms, sill, and Moho (about 12.5 km). The location of shallow dike swarms correspond to airborne dipole magnetic anomalies (Malahoff and Woollard, 1966). Regions of poor resolution reflect limited field data available to inform the model.**

The predicted spatial occurrence of the ocean (left) and specific capacity (right) are shown in Figure 9. The character of the ocean appears correctly located adjacent to the island and increasing at depth to the western and eastern boundary. That said, the ocean at the western edge appears to extend to the bottom of the model domain. In fact, the actual ocean depth in this region is known to be about 6 km thereby revealing the apparent lack of resolution in resistivity and gravity information below this depth to inform the MSOM model. Improvements in predictions for this region of the model would likely require the addition of related properties derived from shipborne and/or ocean bottom geophysical measurements. The specific capacity predictions in this study represent the analog to permeability. In general, there is a trend characterized by low specific capacity at the eastern boundary increasing toward the west. This trend follows from an earlier discussion indicating that the most permeable basaltic rocks are located at the western edge of Lāna‘i. Also, the western dike swarm (between 5 and 8 km depth) is characterized as medium specific capacity, whereas the adjacent region to the east is characterized as low specific capacity.

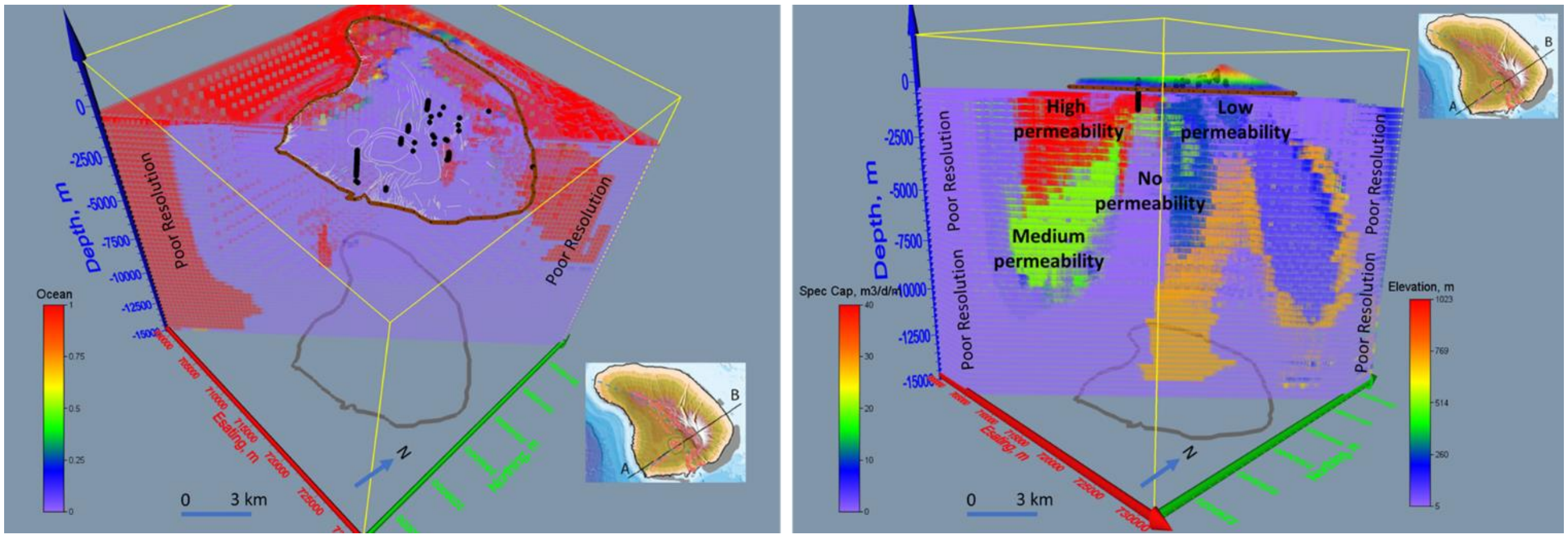

**Figure 9. Predicted features sliced horizontally at sea level and vertically along A-B: (left) 3d ocean character (present//absent), (right) specific capacity (permeability analog) This figure reveals a trend where low specific capacity at the eastern boundary increases toward the west. This trend follows from an earlier discussion indicating that the most permeable basaltic rocks are located at the western edge of Lāna‘i . Well locations are black dots and regions annotated as poor resolution reflect limited field data available to inform the model.**

The predicted spatial distribution of temperature is over the range of 5 C to 758 C as shown in Figure 10. The upper left panel reveals the plan view temperature distribution at sea level with the digital elevation model overlying the island. The deep ocean at the model boundary reveals the coldest temperatures of about 5 C. The upper right panel reveals the plan view temperature distribution at sea level with no digital elevation model. Noteworthy are the cool temperatures near 20 C (blue) in the region of highest elevation and warm temperatures near 65 C (orange) in and around the caldera region. The lower left panel has interpreted regions of downward groundwater recharge from the region of highest elevation and cool temperatures and likely upward convective transport from the 758C source at the Moho through the interpreted dike swarm and sill region of medium specific capacity (surrogate for permeability) toward the warm 65C region of in the vicinity of the deepest well. The region directly below this well and adjacent to the interpreted region of convective transport is predicted to have the no specific capacity (no permeability) thereby supporting plausible convective transport through the more permeable region with conductive heat transfer away from the convective heat transport pathway. The lower right panel identified two geothermal prospects. The first geothermal prospect has predicted temperatures of about 65 C to 108 C and is located between 1 and 2 km depth in the vicinity of the caldera. The second and preferred geothermal prospect has predicted temperatures of about 149 C to 275 C located between 3 to 6 km depths in the vicinity of the western dike swarm and sill.

The predicted spatial distribution of chloride concentration sliced horizontally at sea level and vertically along A-B is shown in Figure 11. The left linear concentration plot spans 23 mg/l (freshwater or no water) to 18000 mg/l (ocean water). The right log chloride concentration plot reveals heterogeneous water types that include freshwater, brackish water, and saline water. Two potential freshwater resource zones are indicated with circles. These freshwater resource zones have similar specific capacity but the zone to the west (closer to the caldera) has elevated temperatures, whereas the zone to the east has cooler temperatures. A conceptual groundwater system is constructed based on chloride concentration, geology, and temperature predictions (figure 12).

### 4.1.3 Feature Clustering

The joint interpretation of information across the MSOM network is undertaken by application of stochastic k-means clustering. In this case, the mode of k-means clusters is interpreted as the most likely and used to aggregate information. The clustering of similar information at various locations characterizes the variability of parameters associated with these Geothermal Stratigraphic Units (GSU). The distribution of GSUs mapped along the cross-section A-B on Lāna‘i reveal five basic groups shown in Figure 13: 1. Ocean (purple). 2. salinized basalt, called salinized basalt (blue); 3. basalt (green); 4. basalt-dike transition, called dike+basalt (orange-red); 5. basalt and dike, called dike (red). Each GSU is characterized by univariate and spatial statistics that describe the physical and chemical properties comprising the geothermal model (Table 3). Other interpreted features include possible underplating and possible Moho at a depth of about 13 km (Lahey et al., 2010).

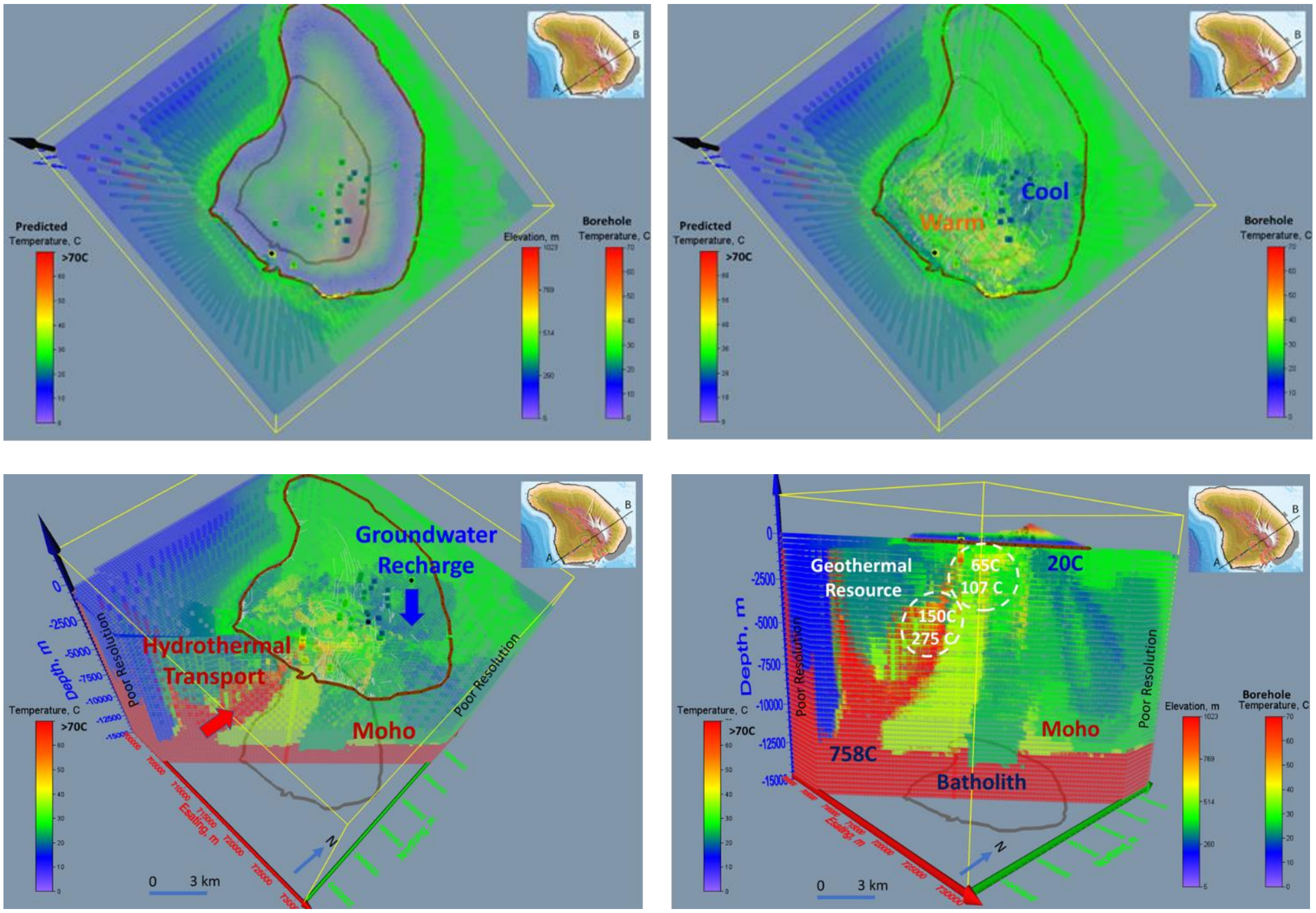

**Figure 10. Predicted temperature over the range of 20 C to 758 C. Upper left: plan view temperature distribution at sea level with the digital elevation model. Upper right: plan view temperature distribution at sea level with no digital elevation model (lower left). Noteworthy are the cool temperatures (blue) attributed to recharge at high elevation and warm temperatures in and around the caldera region. Lower left: interpreted regions of groundwater recharge from the region of highest elevation and hydrothermal transport from the region below the Moho. Lower right: two geothermal prospects; first is about 65 C to 107 C between 1 and 2 km depth in the vicinity of the caldera, second is about 149 C to 275 C between 3 to 6km in the vicinity of the western dike swarm and sill. Regions of poor resolution reflect limited field data available to inform the model.**

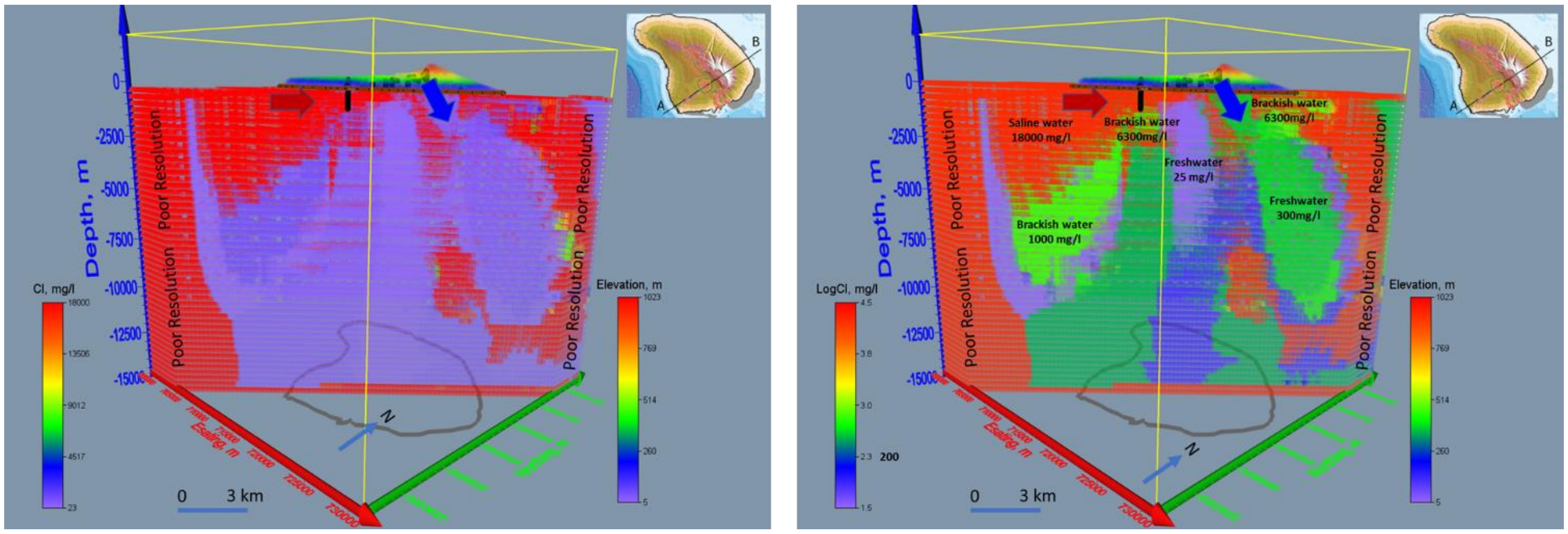

**Figure 11. Predicted chloride concentration sliced horizontally at sea level and vertically along A-B. (left) linear concentration plot (23 mg/l (freshwater or no water) to 18000 mg/l (seawater); (right) log chloride concentration plot showing heterogeneous water types that include freshwater, brackish water, and saline water. Two potential freshwater resource**

**zones are present. The resources have similar permeability but zone to the west (closer to the caldera) has elevated temperatures, whereas the zone to the east has cooler temperatures.**

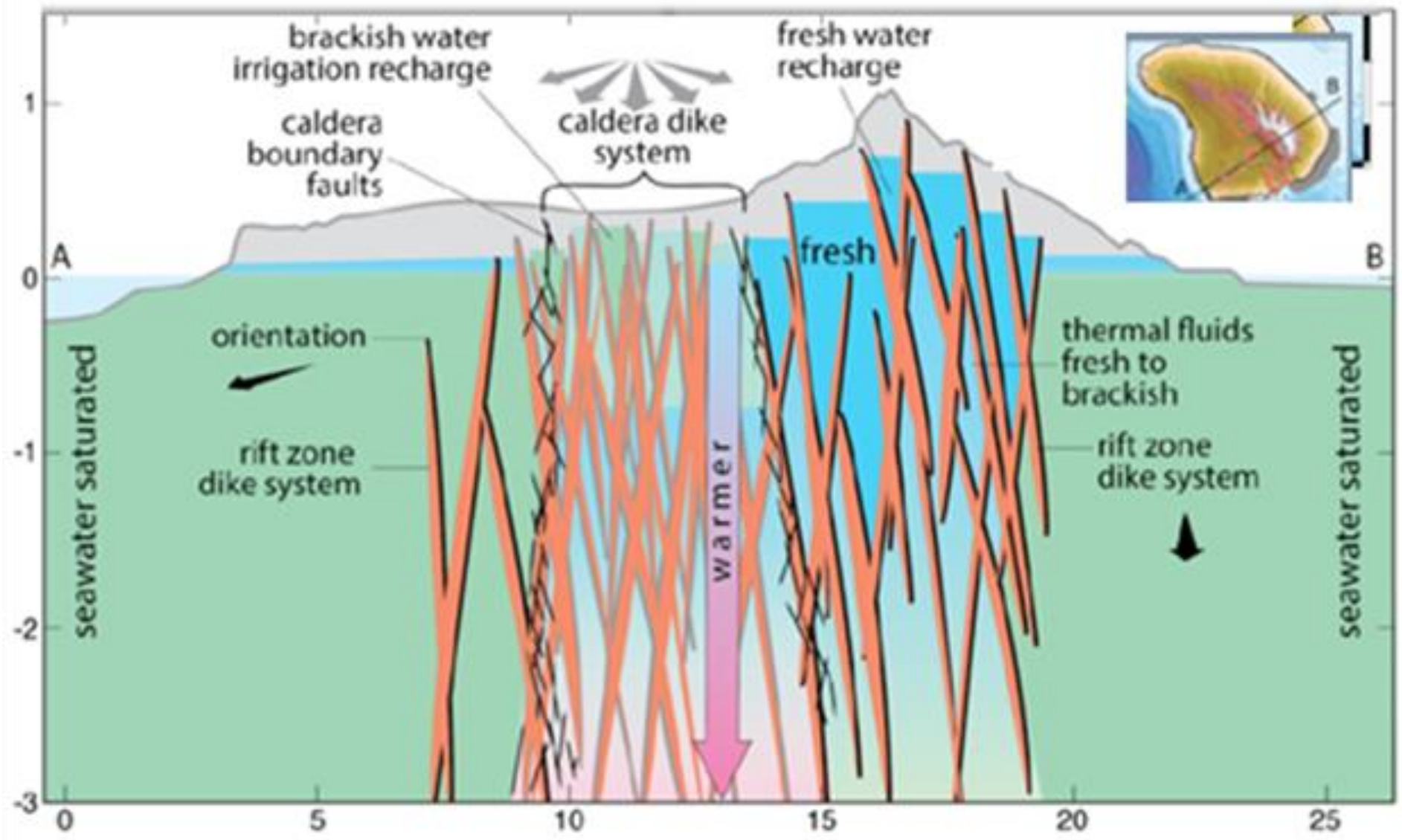

**Figure 12. Conceptual model produced by Lautze and Thomas (unpublished) which agrees with MML predictions of continuous features along cross-section A-B**

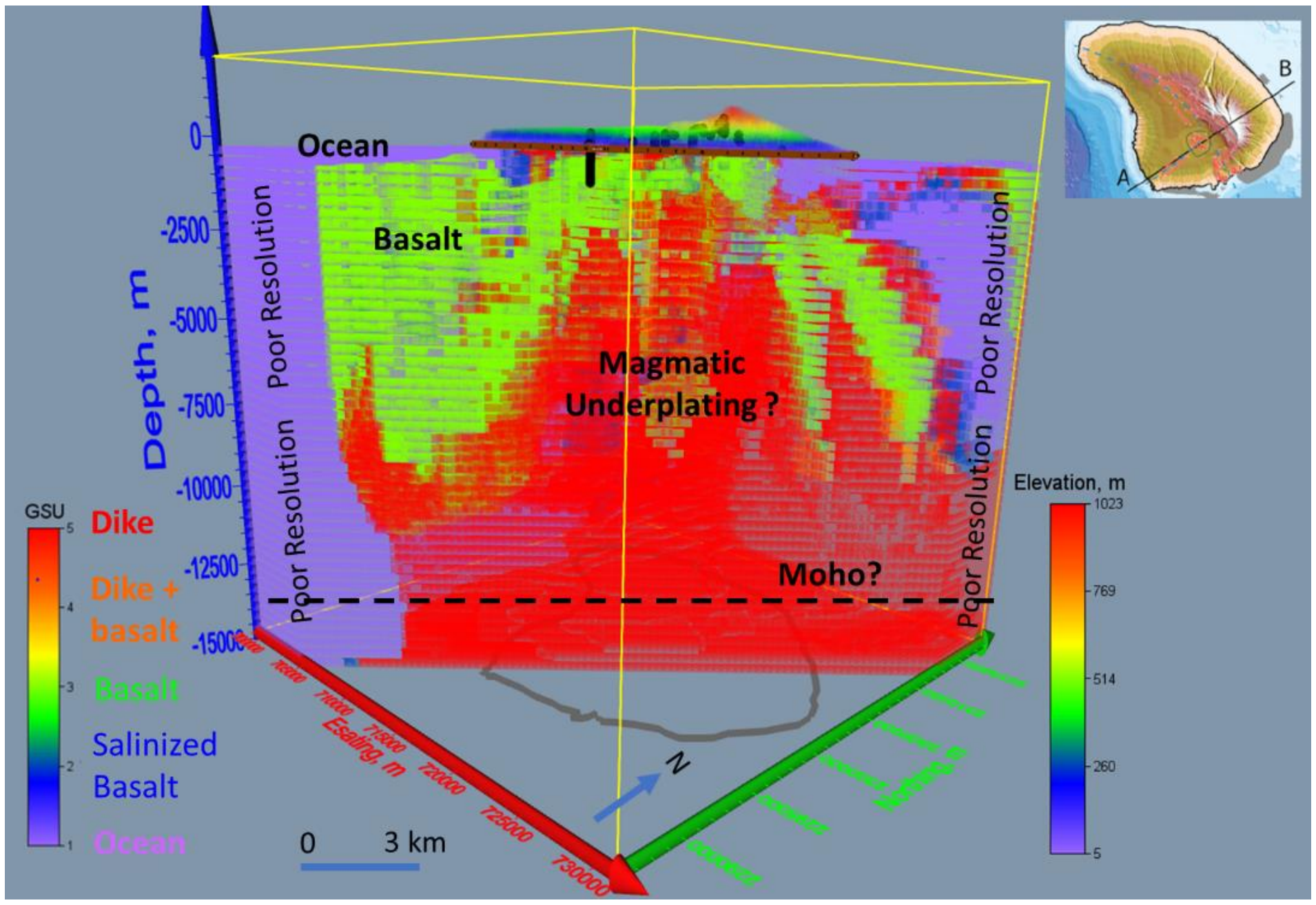

**Figure 13. Predicted 3d Geothermal Stratigraphic Units sliced horizontally at sea level and vertically along A-B reveal five basic groups: 1. ocean (purple), 2. salinized subaerial basalt called salinized basalt (blue), 3. subaerial basalt called basalt (green), 4. submarine basalt-dike transition called dike-basalt (orange), and 5. submarine basalt and dike called dike (red). Each GSU is characterized by univariate and spatial statistics describing all the physical and chemical properties comprising the model (Table 3). Well locations appear as black dots. Other interpreted features include magmatic underplating and Moho (Lahey et al., 2010).**

**Table 3. Summary statistics for selected properties associated with Geothermal Stratigraphic Units. Location; Elev = elevation (m); State variables: Head = hydraulic head (m), cl = chloride concentration (mg/l), Temp = temperature (C); Physical properties: Res = electrical resistivity (ohm-m), SpecCap = specific capacity, (m3/d/m); Aqueous properties: Cond (specific conductance, uM/s, ORP = oxygen reduction potential (mv); Aqueous chemistry: Br = bromide, F = Flouride, Fe = iron, HCO3 = bicarbonate (mg/l), K = potassium, Na = sodium, NO3 = nitrate (mg/l), PO4 = phosphate (mg/l), Si = silica, SO4 = sulfate, Sr = strontium, Ca/Mg = calcium -magnesium ratio.**

| Group | | Summary | Location | | | Water Geology | | | State Variables | | | Physical properties | | | Aqueous properties | | |
|---|---|---|---|---|---|---|---|---|---|---|---|---|---|---|---|---|---|
| GSU | Description | Statistic | Easting, m | Northing, m | Elev, m | Ocean | Dike | Basalt | Head, m | Temp, C | Cl, mg/l | SpeCap, m3/d/m | Density, kg/m3 | Res, ohm-m | Cond, uM/s | ORP, mv | pH |
| 1 | Ocean | minimum | 698886 | 2286203 | -14745 | 1.00 | 0.00 | 0.00 | 0.0 | 4.0 | 16593 | 0.0 | 1020 | 0.1 | 0 | 51.9 | 8.00 |
| | | average | 712723 | 2300699 | -2042 | 1.00 | 0.00 | 0.00 | 0.0 | 58.0 | 17999 | 0.0 | 1325 | 119 | 47961 | 111.3 | 8.00 |
| | | median | 709822 | 2299659 | -319 | 1.00 | 0.00 | 0.00 | 0.0 | 24.4 | 18000 | 0.0 | 1020 | 0 | 48000 | 88.5 | 8.00 |
| | | maximum | 730986 | 2318303 | 2055 | 1.00 | 0.01 | 0.01 | 0.1 | 758.0 | 18000 | 0.0 | 2960 | 3628 | 48000 | 283.1 | 8.00 |
| | | Standard deviation | 10501 | 10412 | 3955 | 0.00 | 0.00 | 0.00 | 0.0 | 157.0 | 21 | 0.0 | 645 | 353 | 1267 | 55.0 | 0.00 |
| | Salanized basalt | | | | | | | | | | | | | | | | |
| 2 | and dikes | minimum | 700086 | 2286803 | -14745 | 0.00 | 0.00 | 0.00 | 0.0 | 19.8 | 164 | 0.0 | 1020 | 0.1 | 0 | 52.1 | 7.95 |
| | | average | 713052 | 2306872 | -1822 | 0.45 | 0.29 | 0.25 | 10.6 | 49.9 | 17611 | 0.0 | 2062 | 60 | 46523 | 112.3 | 8.00 |
| | | median | 711256 | 2309303 | -945 | 0.36 | 0.03 | 0.00 | 0.0 | 26.8 | 18000 | 0.0 | 2515 | 2 | 48000 | 88.5 | 8.00 |
| | | maximum | 730986 | 2318303 | 2055 | 0.98 | 1.00 | 1.00 | 465.1 | 758.0 | 18000 | 7.3 | 2948 | 3818 | 48000 | 283.1 | 8.00 |
| | | Standard deviation | 8740 | 8012 | 3216 | 0.39 | 0.39 | 0.38 | 59.6 | 130.0 | 2102 | 0.3 | 716 | 319 | 7339 | 58.9 | 0.00 |
| 3 | Basalt | minimum | 700756 | 2287403 | -11445 | 0.00 | 0.00 | 0.01 | 0.0 | 18.9 | 23 | 0.9 | 2478 | 4 | 0 | 52.1 | 7.14 |
| | | average | 714600 | 2300283 | -3573 | 0.00 | 0.01 | 0.99 | 42.9 | 29.5 | 5901 | 19.6 | 2717 | 445 | 1365 | 150.9 | 7.87 |
| | | median | 713886 | 2298203 | -2145 | 0.00 | 0.00 | 1.00 | 0.9 | 28.1 | 708 | 20.0 | 2703 | 233 | 166 | 196.2 | 7.90 |
| | | maximum | 730986 | 2318303 | 855 | 0.00 | 0.99 | 1.00 | 484.2 | 73.0 | 18000 | 41.0 | 3011 | 4424 | 48000 | 283.1 | 8.00 |
| | | Standard deviation | 7806 | 7622 | 3165 | 0.00 | 0.10 | 0.10 | 95.4 | 10.1 | 8072 | 13.3 | 96 | 516 | 6207 | 54.8 | 0.13 |
| | Dikes and | | | | | | | | | | | | | | | | |
| 4 | Basalt | minimum | 702786 | 2290103 | -10245 | 0.00 | 0.00 | 0.00 | 0.0 | 18.9 | 31 | 1.4 | 2614 | 3 | 0 | 52.1 | 7.14 |
| | | average | 723764 | 2298945 | -3448 | 0.00 | 0.53 | 0.47 | 263.8 | 28.1 | 7623 | 25.1 | 2769 | 244 | 856 | 151.7 | 7.86 |
| | | median | 724986 | 2298803 | -2445 | 0.00 | 0.57 | 0.43 | 224.0 | 24.5 | 400 | 29.0 | 2747 | 34 | 167 | 196.2 | 7.91 |
| | | maximum | 730986 | 2318303 | 306 | 0.25 | 0.99 | 0.99 | 484.3 | 46.0 | 18000 | 41.0 | 3048 | 4933 | 22370 | 227.4 | 8.00 |
| | | Standard deviation | 5428 | 5197 | 2537 | 0.02 | 0.41 | 0.41 | 172.7 | 9.2 | 8823 | 14.9 | 92 | 603 | 1558 | 58.2 | 0.20 |
| 5 | Dikes | minimum | 698886 | 2286503 | -14745 | 0.00 | 0.00 | 0.00 | 0.0 | 18.9 | 23 | 0.0 | 1020 | 0.1 | 0 | 52.1 | 7.14 |
| | | average | 717025 | 2304238 | -7022 | 0.00 | 0.83 | 0.17 | 229.3 | 182.3 | 6416 | 4.7 | 2768 | 748 | 14322 | 153.7 | 7.79 |
| | | median | 717186 | 2304259 | -7245 | 0.00 | 1.00 | 0.00 | 246.3 | 26.8 | 543 | 0.0 | 2807 | 237 | 2036 | 135.3 | 8.00 |
| | | maximum | 730986 | 2318303 | 2055 | 0.80 | 1.00 | 1.00 | 485.6 | 758.0 | 18000 | 41.0 | 3055 | 8066 | 48000 | 283.1 | 8.00 |
| | | Standard deviation | 7819 | 8007 | 4903 | 0.01 | 0.37 | 0.37 | 179.8 | 297.8 | 8376 | 8.9 | 262 | 1166 | 20843 | 58.2 | 0.34 |

| Group | | Summary | Location | | | Water Geology | | | Aqueous chemistry | | | | | | | | | | | |
|---|---|---|---|---|---|---|---|---|---|---|---|---|---|---|---|---|---|---|---|---|
| GSU | Description | Statistic | Easting, m | Northing, m | Elev, m | Ocean | Dike | Basalt | HCO3, mg/l | NO3, mg/l | PO4, mg/l | SO4, mg/l | Br, mg/l | F, mg/l | Fe, mg/l | K, mg/l | Na, mg/l | Si, mg/l | Sr, mg/l | Ca/Mg |
| 1 | Ocean | minimum | 698886 | 2286203 | -14745 | 1.00 | 0.00 | 0.00 | 139.0 | 0.70 | 0.10 | 2.65 | 60.97 | 1.38 | 0.02 | 357.1 | 9200 | 1.0 | 0.02 | 0.31 |
| | | average | 712723 | 2300699 | -2042 | 1.00 | 0.00 | 0.00 | 140.0 | 0.70 | 0.10 | 2.65 | 64.00 | 1.40 | 0.02 | 380.0 | 10000 | 1.0 | 1.04 | 0.31 |
| | | median | 709822 | 2299659 | -319 | 1.00 | 0.00 | 0.00 | 140.0 | 0.70 | 0.10 | 2.65 | 64.00 | 1.40 | 0.02 | 380.0 | 10000 | 1.0 | 0.38 | 0.31 |
| | | maximum | 730986 | 2318303 | 2055 | 1.00 | 0.01 | 0.01 | 140.0 | 0.70 | 0.12 | 2.67 | 64.00 | 1.40 | 0.02 | 380.0 | 10000 | 1.1 | 2.66 | 0.32 |
| | | Standard deviation | 10501 | 10412 | 3955 | 0.00 | 0.00 | 0.00 | 0.0 | 0.00 | 0.00 | 0.00 | 0.04 | 0.00 | 0.00 | 0.3 | 11 | 0.0 | 1.11 | 0.00 |
| | Salanized basalt | | | | | | | | | | | | | | | | | | | |
| 2 | and dikes | minimum | 700086 | 2286803 | -14745 | 0.00 | 0.00 | 0.00 | 103.6 | 0.62 | 0.10 | 2.65 | 7.51 | 0.59 | 0.01 | 27.4 | 311 | 1.0 | 0.04 | 0.31 |
| | | average | 713052 | 2306872 | -1822 | 0.45 | 0.29 | 0.25 | 139.8 | 0.70 | 0.10 | 2.65 | 63.19 | 1.40 | 0.02 | 376.8 | 9896 | 1.0 | 0.74 | 0.32 |
| | | median | 711256 | 2309303 | -945 | 0.36 | 0.03 | 0.00 | 140.0 | 0.70 | 0.10 | 2.65 | 64.00 | 1.40 | 0.02 | 380.0 | 10000 | 1.0 | 0.34 | 0.31 |
| | | maximum | 730986 | 2318303 | 2055 | 0.98 | 1.00 | 1.00 | 140.0 | 0.72 | 1.51 | 3.81 | 64.00 | 1.40 | 0.02 | 380.0 | 10000 | 27.9 | 2.60 | 0.67 |
| | | Standard deviation | 8740 | 8012 | 3216 | 0.39 | 0.39 | 0.38 | 1.7 | 0.00 | 0.05 | 0.05 | 4.70 | 0.04 | 0.00 | 22.8 | 713 | 0.9 | 0.84 | 0.01 |
| 3 | Basalt | minimum | 700756 | 2287403 | -11445 | 0.00 | 0.00 | 0.01 | 52.5 | 0.48 | 0.00 | 2.65 | 0.06 | 0.00 | 0.00 | 1.9 | 16 | 1.0 | 0.04 | 0.31 |
| | | average | 714600 | 2300283 | -3573 | 0.00 | 0.01 | 0.99 | 84.9 | 1.69 | 2.25 | 27.11 | 3.78 | 0.11 | 0.17 | 22.6 | 526 | 1078 | 0.21 | 1.05 |
| | | median | 713886 | 2298203 | -2145 | 0.00 | 0.00 | 1.00 | 83.0 | 0.64 | 2.58 | 6.40 | 0.14 | 0.04 | 0.00 | 2.7 | 23 | 880 | 0.06 | 1.13 |
| | | maximum | 730986 | 2318303 | 855 | 0.00 | 0.99 | 1.00 | 146.4 | 4.24 | 4.84 | 151.20 | 64.00 | 1.40 | 0.78 | 380.0 | 10000 | 1650 | 0.74 | 1.22 |
| | | Standard deviation | 7806 | 7622 | 3165 | 0.00 | 0.10 | 0.10 | 25.9 | 1.63 | 2.12 | 33.57 | 13.69 | 0.30 | 0.31 | 80.8 | 2135 | 418 | 0.17 | 0.21 |
| | Dikes and | | | | | | | | | | | | | | | | | | | |
| 4 | Basalt | minimum | 702786 | 2290103 | -10245 | 0.00 | 0.00 | 0.00 | 52.5 | 0.48 | 0.00 | 2.65 | 0.06 | 0.00 | 0.00 | 1.9 | 16 | 1.0 | 0.04 | 0.31 |
| | | average | 723764 | 2298945 | -3448 | 0.00 | 0.53 | 0.47 | 71.5 | 0.90 | 2.05 | 11.49 | 0.65 | 0.04 | 0.06 | 4.7 | 74 | 1037 | 0.23 | 1.08 |
| | | median | 724986 | 2298803 | -2445 | 0.00 | 0.57 | 0.43 | 83.0 | 0.64 | 2.58 | 7.20 | 0.14 | 0.04 | 0.00 | 2.1 | 22 | 1130 | 0.38 | 1.13 |
| | | maximum | 730986 | 2318303 | 306 | 0.25 | 0.99 | 0.99 | 142.8 | 4.26 | 3.87 | 76.00 | 64.00 | 1.40 | 0.78 | 380.0 | 10000 | 1650 | 0.74 | 1.22 |
| | | Standard deviation | 5428 | 5197 | 2537 | 0.02 | 0.41 | 0.41 | 16.9 | 0.97 | 1.76 | 18.53 | 4.49 | 0.11 | 0.21 | 26.0 | 677 | 259.4 | 0.16 | 0.15 |
| 5 | Dikes | minimum | 698886 | 2286503 | -14745 | 0.00 | 0.00 | 0.00 | 52.5 | 0.48 | 0.00 | 2.65 | 0.06 | 0.00 | 0.00 | 1.9 | 16 | 1.0 | 0.04 | 0.31 |
| | | average | 717025 | 2304238 | -7022 | 0.00 | 0.83 | 0.17 | 122.2 | 2.19 | 1.64 | 26.37 | 24.25 | 0.71 | 0.07 | 186.0 | 4842 | 645 | 0.57 | 0.66 |
| | | median | 717186 | 2304259 | -7245 | 0.00 | 1.00 | 0.00 | 140.0 | 0.70 | 0.10 | 6.40 | 1.88 | 0.13 | 0.02 | 12.7 | 127 | 842 | 0.38 | 0.84 |
| | | maximum | 730986 | 2318303 | 2055 | 0.80 | 1.00 | 1.00 | 146.4 | 7.59 | 5.81 | 151.20 | 64.00 | 1.40 | 0.78 | 380.0 | 10000 | 1650 | 2.61 | 1.22 |
| | | Standard deviation | 7819 | 8007 | 4903 | 0.01 | 0.37 | 0.37 | 26.5 | 2.56 | 2.31 | 38.14 | 29.35 | 0.67 | 0.20 | 186.8 | 4961 | 654 | 0.79 | 0.35 |

## 4.2 Island of Hawai'i

### 4.2.1 Feature Selection

The feature selection process undertaken using learn heuristics (wrapper method) is constrained by the state variable temperature (Jerran et al., 2019, Stolper et al., 2009, Sorey and Colvard, 1994; Puna Geothermal Venture, 1991, State of Hawai'i, 1990, Kihara et al., 1977; Kingston Reynolds Thom & Allardice Ltd., 1976) shown in the purple box. Informative features appearing in the green box are determined to be associated with geology (e.g., various fault systems and basalt, oceanic crust, underplating and mantle layering) and geophysical properties (e.g., density, resistivity, p-wave and s-wave velocity, and earthquake magnitudes). The following sections cover three parts: 1. Local Assimilation and Prediction, 2. Regional Feature Assimilation and Prediction, and 3. Island-Wide Assimilation and Prediction.

Of the original predictor variables available for consideration (N = 36), the learn-heuristic based feature-selection process identified an optimal set of global variables (N=21) suitable for predicting features beneath the island of Hawai'i (Table 4). These numeric and categorical features are used in constructing, training, and testing the multimodal machine learning models. Numeric features that appear informative to the state variable (temperature) include location: easting, northing, elevation; and geophysical properties: density, resistivity, p-wave velocity, s-wave velocity, and earthquake magnitude. Categorical features that appear informative to the to the state variable (temperature) include geology: Hilina, Kahuku, Kaoiki, Kealakehua, Kohala, and Koae fault systems: and Kilauea and Mauna Loa rifts (Lipman et al, 1990, Wolfe and Morris, 1996; Trusdell et al., 2006); and basalt, oceanic crust, and mantle layers (Leahy et al., 2010); Unlike the Lānaʻi study, there are no aqueous properties, aqueous chemistry, or stable isotopes included during phase 1 of this study.

**Table 4. Summary of features deemed informative using the learn heuristics approach constrained to the state variables temperature (C) are in three categories: location (Easting, Northing, Elevation), geology (Hilina, Kahuku, Kaoiki, Kealakehua, Kilauea, Koae, Kohala, and Mauna Loa fault systems (Lipman et al, 1990, Wolfe and Morris, 1996; Trusdell et al., 2006); basalt, oceanic crust, and mantle layers (Leahy et al., 2010) and geophysical properties: density, resistivity, p-wave velocity, s-wave velocity, and earthquake magnitude. Unlike the Lānaʻi study, there are no aqueous properties, aqueous chemistry, or stable isotopes included during phase 1.**

Constraints on learn heuristics

Informative features

| | | Type | | Observation | | Support | Location | | |
|---|---|---|---|---|---|---|---|---|---|
| Category | Features | Determinitic | Stochastic | Measured | Derived | Point-1D | Borehole | Surface | Source |
| Location | Easting, m | X | | X | | X | | | |
| | Northing, m | X | | X | | X | | | |
| | Elevation, m | X | | X | | X | | | |
| State variable | Temperature, C | X | | X | X | X | | | Lautze et al., 2020; Putrikia, 1997 |
| Geology | Hilina Fault | X | | X | | X | | | Lipman et al., 1990; |
| | Kahuku Fault | X | | X | | X | | | Wolfe and Morris, 1996; |
| | Kaoiki Fault | X | | X | | X | | | Trusdell et al., 2006 |
| | Kealakekua Fault | X | | X | | X | | | |
| | | X | | X | | X | | | |
| | Koae Fault | X | | X | | X | | | |
| | Kohala Fault | X | | X | | X | | | |
| | Kilauea Rift | X | | X | | X | | | |
| | Mauna Loa Rift | | | | | | | | |
| | Basalt Layer | X | | | X | X | | | Leahy et al., 2010 |
| | Oceanic Crust | X | | | X | X | | | Leahy et al., 2010 |
| | Under plating | X | | | X | X | | | Leahy et al., 2010 |
| | Mantle | X | | | X | X | | | Leahy et al., 2010 |
| Physical properties | Density, kg/m3 | X | | | X | X | | | Lautze et al., 2020 |
| | Resistivity, ohm-m | X | X | | X | X | X | X | Lautze et al., 2020 |
| | P-wave velocity, km/s | X | | | X | X | | X | Lautze et al., 2020 |
| | S-wave velocity, km/s | X | | | X | X | | X | Lautze et al., 2020 |
| | Vp/Vs | X | | | | | | | |
| | Earthquake, magnitude | X | | | | X | | X | Lautze, et al., 2020 |

### 4.2.2 Feature Prediction

Prior to training, a normalization procedure is applied to elevations (spanning depths from 4.1 to -40 km) producing a vector of values spanning the range of 0 to 1. The remaining numeric features are normalized by their data variance and randomly assigned (presenting the input vectors to the map sequentially using a randomly sorted database) as an initial set of map weight vectors. Application of the MSOM network to training data is done using a single fixed number of neurons and topological relations which differ for each model appearing the following sections: local assimilation and feature prediction (simple), regional assimilation and feature prediction (intermediate in complexity) and island-wide assimilation and feature prediction briefly described next.

#### 4.2.2.1 Local Assimilation and Feature Prediction

In this section, we present local (separate) feature assimilation and prediction results for the Saddle Rd. and Lower East Rift Zone areas. The aim is to assess the quality of sparse measurements to inform local predictions. The objectives are to: (1) predict Saddle Rd resistivity at MT/AMT stations (blue) & borehole (yellow), and (2) predict temperature at Lower east rift zone (right) boreholes (yellow).

At Saddle Rd, the resistivity model is developed using location (Easting, Northing, Elevation), density, and resistivity (deterministic layered earth inversions to 40 km depth by personnel at the University of Hawai'i and stochastic layered earth inversions to 3 km depth by personnel at the Pacific Northwest National Laboratory). The MSOM shape (26 rows by 20 columns) is a toroid (wraps from top to bottom and side to side) with hexagonal neurons. Training of the map is conducted using both rough and fine phases. The rough training phase involved 20 iterations using a Gaussian neighborhood with an initial and final radius of 33 units and 9 units; and the fine training involved 400 iterations using a Gaussian neighborhood with an initial and final radius of 9 units and 1 unit. The initial and final learning rates of 0.5 and 0.05 decay linearly down to $10^{-5}$, and the Gaussian neighborhood function decreases exponentially from a decay rate of $10^{-1}$ iteration to $10^{-3}$, providing reasonable convergence evidenced by similarity in their low quantization ($q_e$=0.21) and topographic ($t_e$=0.145) errors.

The log-resistivity prediction results for station 12PT are presented in Figure 14. This figure provides: (a) the location map highlighting resistivity station 12PT, (b) observed (deterministically inverted 1D layer earth resistivity values) and predicted resistivity values (minimum, median, and maximum values determined using 30 trials and a leave one out cross-validation strategy). (c) plot of the deterministic log resistivity (LogRho) versus median stochastic log resistivity (LogQ50), and (d) prominent resistive and conductive features from the surface to 40 km based on the deterministic inversion. Note that there is a slight bias in the prediction of resistivity associated with conductive zones suggesting the potential need to increase the number of model trials, increase the network size, and/or add additional features to better constrain the prediction. The strong correlation coefficient (0.92) between LogRho (University of Hawai'i resistivity values to depths of 40 km) and median LogQ50 (Pacific Northwest National Lab resistivity values to 3 km) supports the ability to assimilate shallow higher frequency resistivity information with deeper lower frequency resistivity information (Figure 14c).

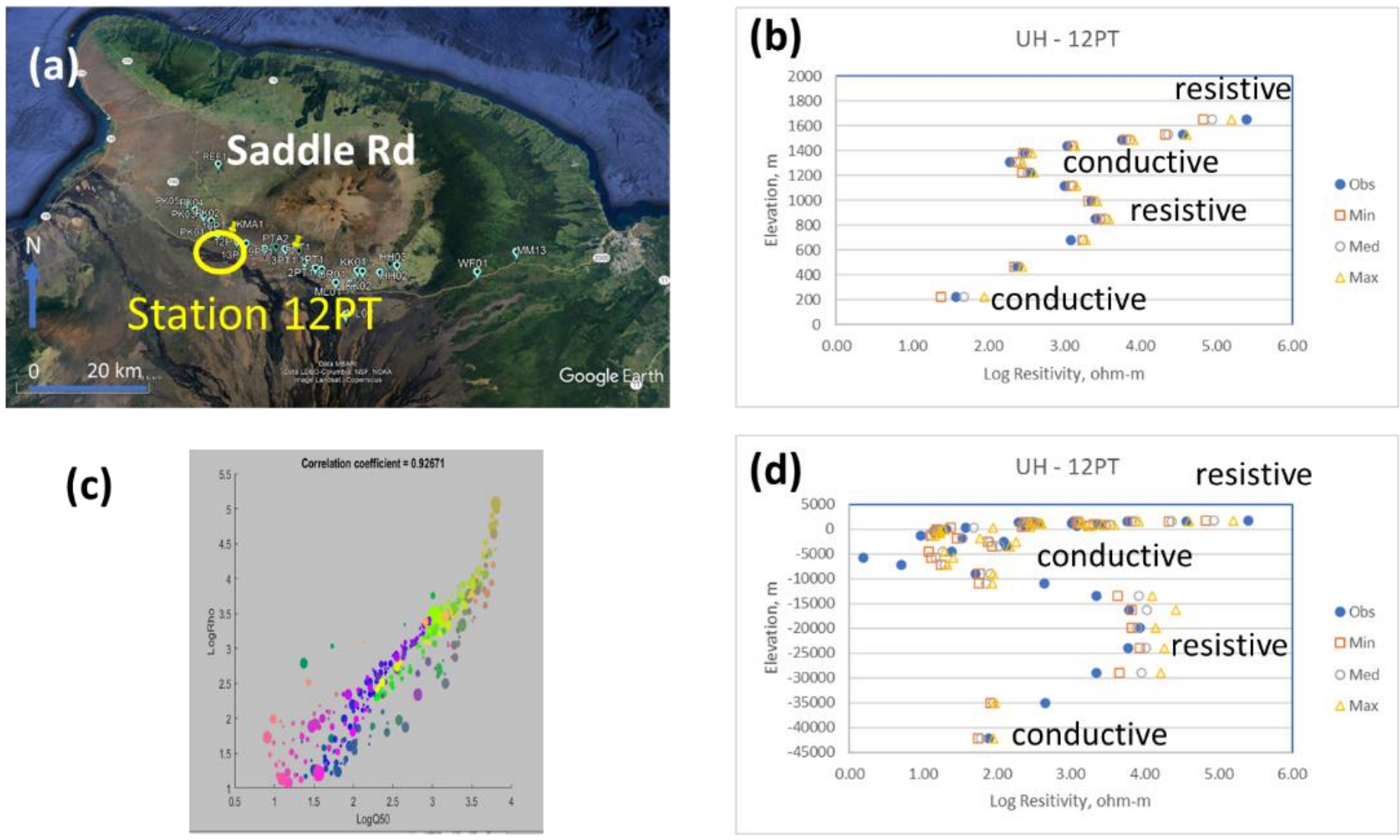

**Figure 14. Predicted Saddle Rd 1D resistivity features: (a) The location map highlighting resistivity station 12PT. (b) Observed and predicted resistivity values at station 12PT. The observed values are plotted with minimum, median, and maximum values determined using 30 Trials and a leave one out cross-validation strategy. There is slight bias in the conductive zones suggesting the need to add additional features as prediction constraints. (c) Plot of the deterministic log resistivity (LogRho) versus median stochastic log resistivity (LogQ50), and (d) Prominent resistive and conductive features from the surface to 40 km based on the deterministic inversion.**

The Lower East Rift Zone temperature model is developed using location (Easting, Northing, and Elevation) and sparse temperatures from geothermal wells across the area. In this regard, the MML workflow is used as a nonlinear multivariate 3d interpolator. The neural map shape (22 rows by 16 columns) is a toroid (wraps from top to bottom and side to side) with hexagonal neurons. Training of the map was conducted using both rough and fine phases. The rough training phase involved 20 iterations using a Gaussian neighborhood with an initial and final radius of 28 units and 7 units; and the fine training involved 400 iterations using a Gaussian neighborhood with an initial and final radius of 7 units and 1 unit. The initial and final learning rates of 0.5 and 0.05 decayed linearly down to $10^{-5}$, and the Gaussian neighborhood function decreased exponentially from a decay rate of $10^{-1}$ iteration to $10^{-3}$, providing reasonable convergence evidenced by similarity in their low quantization ($q_e$=0.10) and topographic ($t_e$=0.14) errors.

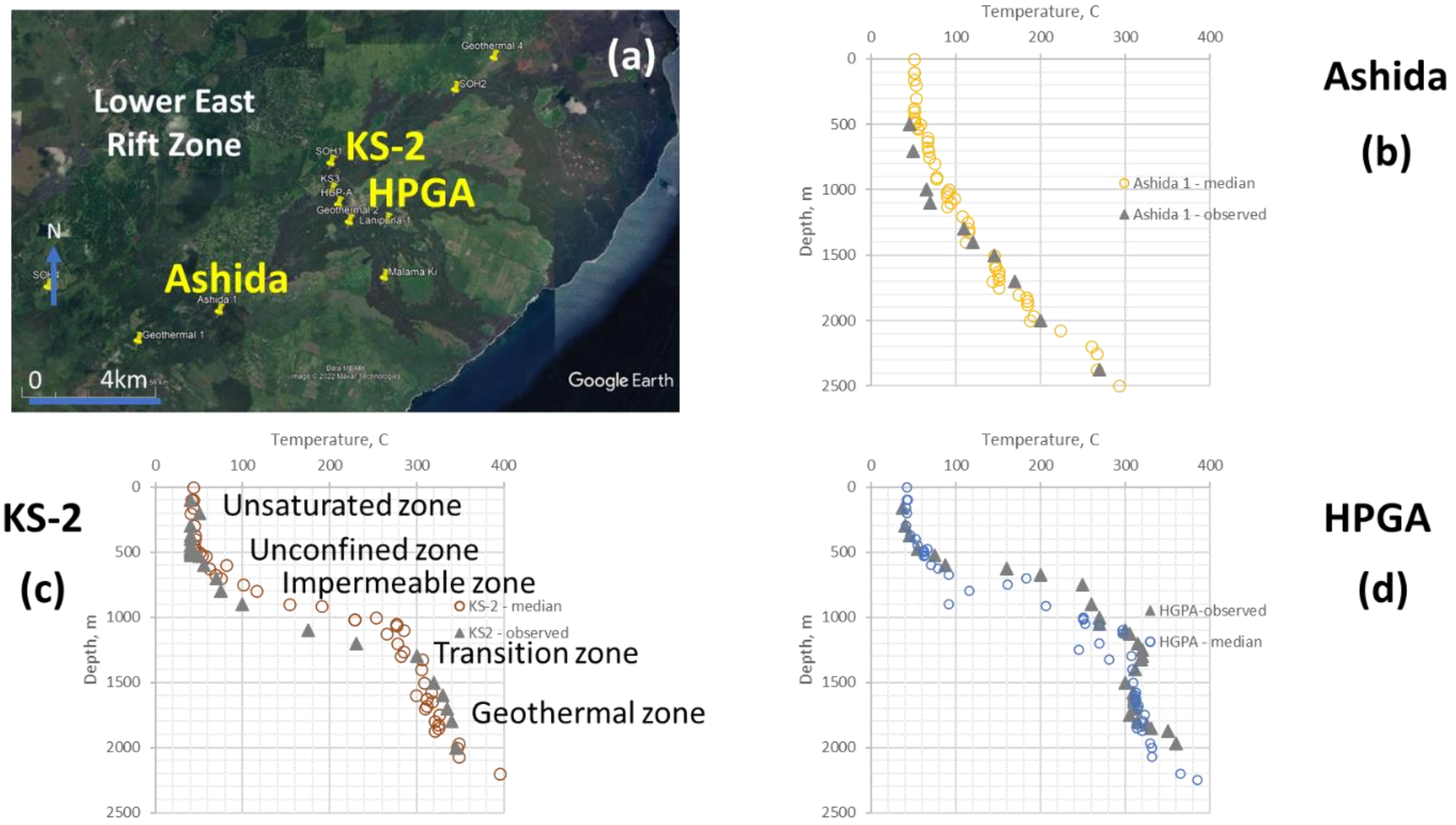

**Figure 15. Predicted Lower East Rift Zone temperature predictions: (a) Location map for three geothermal wells (yellow): Ashida, KS-2, and HPGA; (b) observed and median prediction temperatures at the Ashida well; (c) observed and median prediction temperatures at the KS-2 well; (b) observed and median prediction temperatures at the HPGA well. The observed temperatures appear as triangles and median predictions appear as circles. The median prediction values are computed following 30 random trials using a leave one-out cross-validation strategy. The KS-2 temperature profile is interpreted as having five zones: unsaturated, unconfined, impermeable, transition, and geothermal. Similar zonation can be identified in the HPGA well but not in the Ashida well.**

Comparative temperature profiles are presented for three geothermal wells located in the Lower East Rift Zone (Figure 15). The location map (Figure 15a) shows their relative proximity to each other whereas the other figures present the measured and predicted temperature profiles for Ashida (b), KS-2 (c) , and HPGA (d) geothermal wells. The observed temperatures appear as triangles and median predictions appear as circles. The median prediction values are computed following 30 trials using a leave one-out cross-validation strategy. In general, there is visual correspondence among observations and median prediction values at these geothermal wells. Note that these plots appear in order of increasing thermal complexity despite their proximity (less than 3 km separation). Of these profiles, the KS-2 temperature plot characterizes well-defined hydrologic zones that include unsaturated, unconfined, impermeable, transition, and geothermal. These transitions zones could potentially provide an additional set of categorical constraints in future studies with the aim of identifying hidden groundwater resources. Other available categorical measurements available to constrain feature predictions could consider subsurface geologic information, such as subaerial, shallow marine, transition, and deep marine environments (Campbell and Gardner, 1981).

4.2.2.2 Regional Assimilation and Prediction

In this section, the regional feature assimilation and prediction results are presented when combining data from the Saddle Rd., Kilauea, and Lower East Rift Zone area. The aim is to identify potential 2D hidden geothermal resources, 2D lithospheric flexure under volcanic loading, and 2D geologic layering. The objectives are twofold: (1) predict geophysical (numeric) features, e.g., temperature, density, resistivity, Vp, and Vp/Vs ratio; and (2) predict geologic (categorical) features, e.g velocity layers basalt, crust, underplating, and mantle.

The regional model developed herein incorporates information from these numeric and categorical features. The map shape (72 rows by 66 columns) is a toroid (wraps from top to bottom and side to side) with hexagonal neurons. Training of the map was conducted using both rough and fine phases. The rough training phase involved 20 iterations using a Gaussian neighborhood with an initial and final radius of 98 units and 25 units; and the fine training involved 400 iterations using a Gaussian neighborhood with an initial and final radius of 25 units and 1 unit. The initial and final learning rates of 0.5 and 0.05 decayed linearly down to $10^{-5}$, and the Gaussian neighborhood function decreased exponentially from a decay rate of $10^{-1}$ iteration to $10^{-3}$, providing reasonable convergence evidenced by their reasonably low quantization ($q_e$=0.053) and topographic ($t_e$=0.34) errors. Given that all three split sets (folds 1, 2, and 3) have similar error vectors, future studies might consider increasing the network size to reduce the topographic error vector (the proportion of all data vectors for which first and second BMUs are not adjacent units).to be of the same magnitude as the quantization error (average distance between each data vector and its BMU and is a measure of map resolution). This approach has been shown by Friedel et al. (2020) to reduce uncertainty and bias thereby improving feature predictions.

The component planes plot, shown in Figure 16, presents the converged weights for the first 20% subsample (fold-1) of records across the MSOM network by numeric (location: easting, northing, elevation), state variable (temperature), geophysical properties (density, log resistivity, vp, vp/vs, and earthquake magnitude) and categorical (fault systems: Hilina, Kahuku, Kaoiki, Kealakekua, Koae, Kohala; rfits: Kilauea and Mauna Loa; and geologic layers: basalt, oceanic crust, underplating, and mantle). Similar pattern and colors imply high spatial correlation, e.g., temperature and vp (p-wave velocity), whereas same pattern with opposite colors imply high negative spatial correlation, e.g., temperature and elevation. Variations among feature weights with the state variable is preferrable to highly correlated features which would increase the nonuniqueness in predictions and therefore uncertainty.

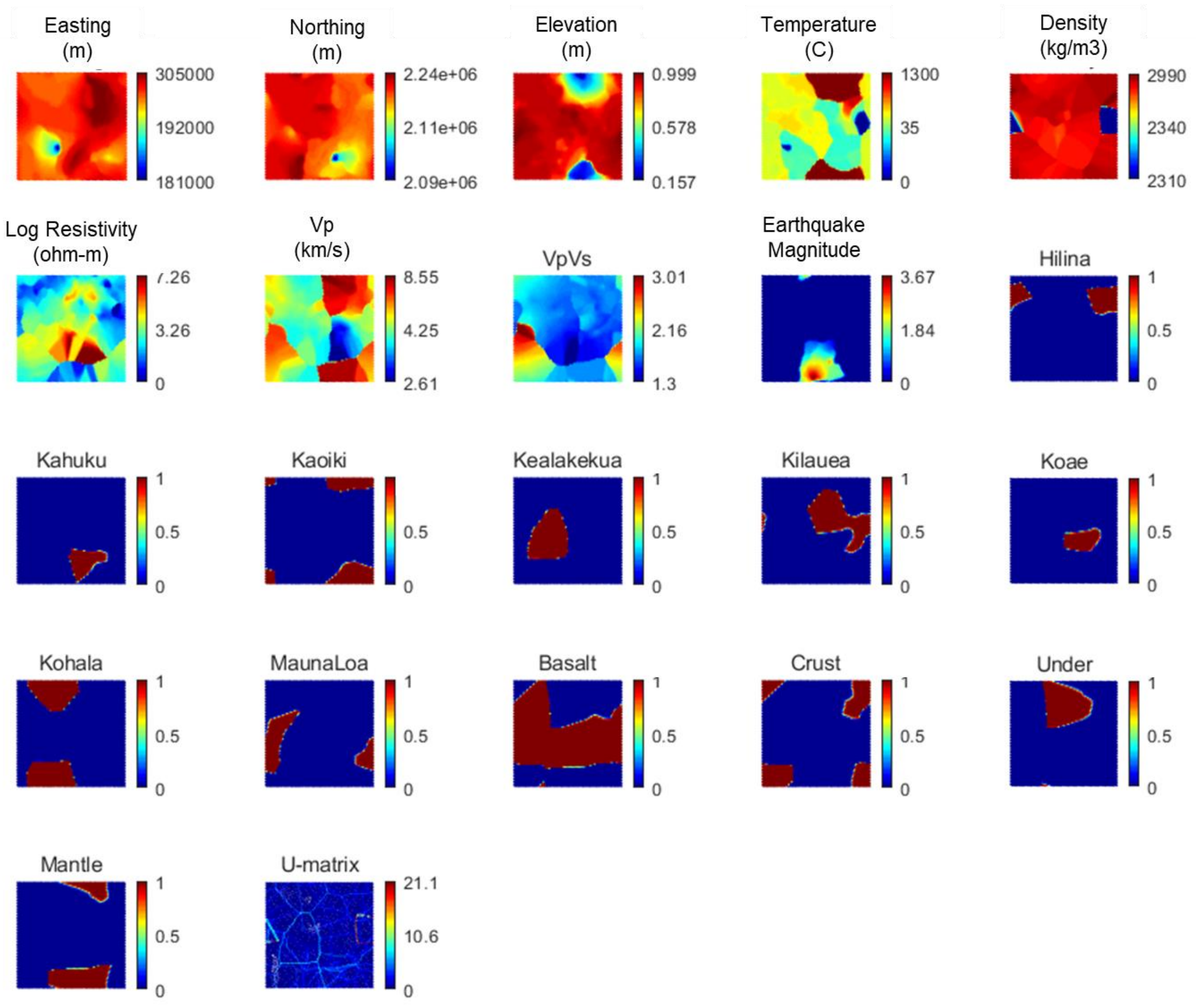

**Figure 16. Component planes plot (Vesanto, 1999) shows the converged weights first 20% subsample (fold-1) of records across the MSOM network by numeric (location: easting, northing, elevation), state variable (temperature), geophysical properties(density, log resistivity, vp, vp/vs, and earthquake magnitude) and categorical (fault systems: Hilina, Kahuku, Kaoiki, Kealakekua, Kilauea, Koae, Kohala, Mauna Loa; and geologic layers: basalt, oceanic crust, underplating, and mantle). Similar pattern and colors imply high spatial correlation, e.g. temperature and vp (p-wave velocity), whereas same pattern with opposite colors imply high negative spatial correlation, e.g. temperature and elevation. Variations among feature weights with the state variable is preferrable to highly correlated features which would increase the nonuniqueness in predictions and therefore uncertainty.**

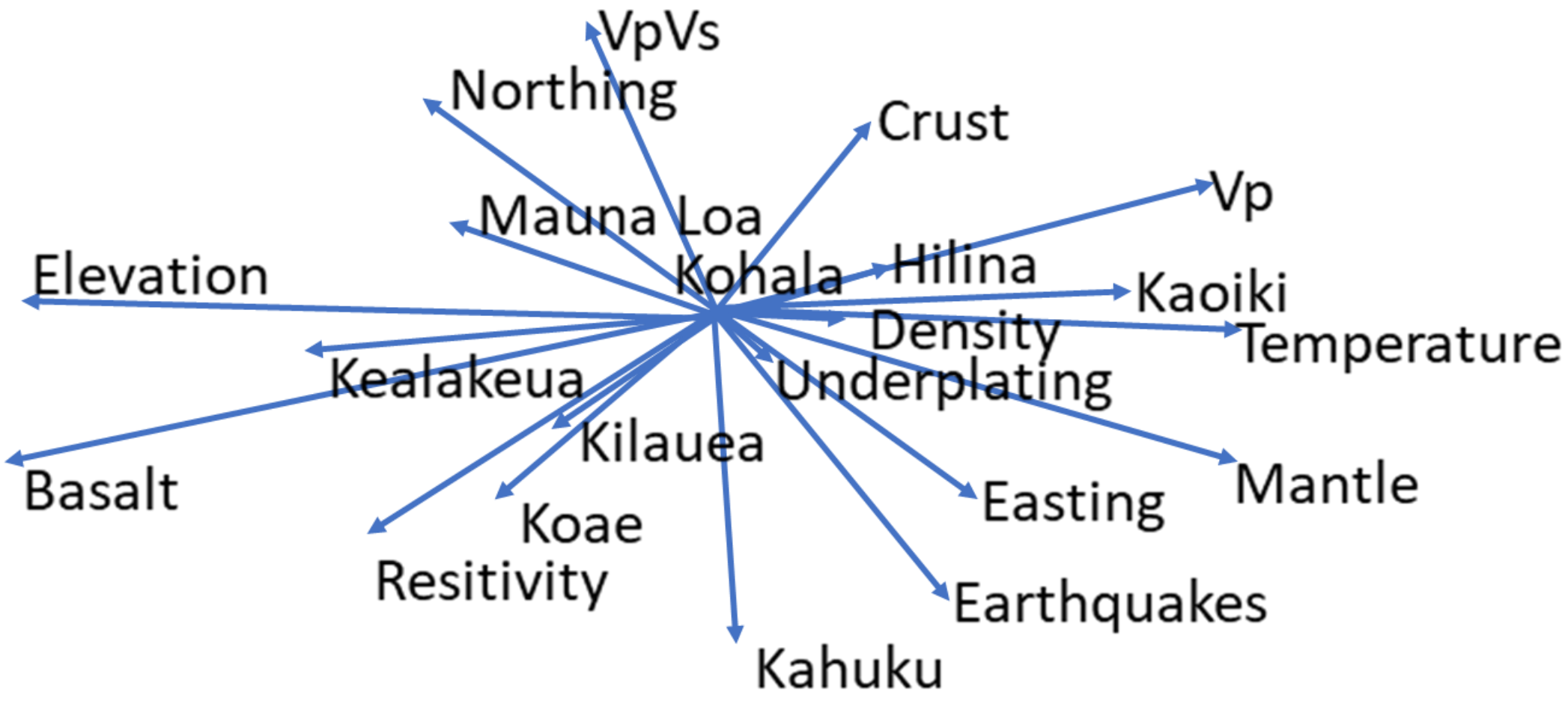

**Figure 17. Principal component planes projection of median feature weight vectors (Vesanto, 1999) following the fold-1 split. Features on opposite sides of the plat are anticorrelated, e.g. elevation and temperature and basalt and vp, whereas the features that group together are more likely to be positively related, e.g. earthquakes and easting, vpvs and northing, density and underplating, and Koae and resistivity. A preferred model will have features distributed around the origin. Lastly, features with fewer data will plot closer to the origin whereas those with more data will plot toward the outside. The magnitude of these vectors (number of records) in part dependent on the randomly shuffled split.**

Testing and validation of the trained MSOM model are reflective of the three split sets previously discussed, where each k-fold represents a random sampling of shuffled records split (80/20) from original data set. Summary statistics following independent testing when predicting temperature, density, resistivity, p-wave velocity, and s-wave velocity are provided in Table 5. For the sake of brevity, we present cross-plots and their associated $R^2$ values for temperature for each fold in figure 18 and Table 6. In these figures, the independent observations appear along the x-axis and predicted observations appear along the y-axis. The high $R^2$ values provided in the table for each fold suggest that the model generalizes well when presented with independent data. Note that in contrast to the Lānaʻi study, the metrics for temperature predictions are reflective of those measured in the upper 3 km and those applied as a linear boundary condition (based on regressed temperature values determined using thermobarometry, Putrika, K., 1997) at depths from 20 to 40 km.

In addition to statistical summaries for the independent testing of numeric features, we provide tables of kappa statistics for the independent testing of categorical features (Table 7). These tables present kappa statistics that reveal a reasonably high level of independent model performance when predicting categorical k-fold layer features such basalt, oceanic crust, magmatic underplating and mantle for three different folds (subsample splits). In general, the independent model layer predictions range from moderate to almost perfect across the 3 independent split sets (folds).

**Table 5: Summary statistics following independent testing of selected features predicted using the trained machine learning network. Each k-fold reflects a randomly sampling of shuffled records split (80/20) from original data set.**

| Fold-1 | Statistic | | | | | | | | | |
|---|---|---|---|---|---|---|---|---|---|---|
| Feature | N | N* | Mean | SE Mean | StDev | Minimum | Q1 | Median | Q3 | Maximum |
| Temperature Observed, C | 3960 | 14235 | 232 | 6.32 | 398 | 17.2 | 63.9 | 71.5 | 98.8 | 1301 |
| Temperature Predicted, C | 18195 | 0 | 188 | 2.58 | 348 | 14.9 | 59.1 | 72.9 | 101.7 | 1299 |
| Density Observed, kg/m3 | 223 | 17972 | 2690 | 4.92 | 73.5 | 2499 | 2637 | 2678 | 2747 | 2993 |
| Density Predicted, kg/m3 | 18195 | 0 | 2650 | 0.86 | 116 | 2314 | 2628 | 2659 | 2718 | 3012 |
| Log Resisitivity Observed, ohm-m | 412 | 17783 | 2.54 | 0.06 | 1.17 | -0.80 | 1.65 | 2.43 | 3.37 | 6.87 |
| Log Resisitivity Predicted, ohm-m | 18195 | 0 | 2.65 | 0.01 | 1.14 | -0.68 | 2.17 | 2.55 | 3.05 | 7.64 |
| Vp Observed, km/s | 3453 | 14742 | 4.90 | 0.02 | 1.24 | 2.69 | 4.04 | 4.43 | 5.25 | 8.60 |
| Vp Predicted, km/s | 18195 | 0 | 4.79 | 0.01 | 1.30 | 2.67 | 4.04 | 4.33 | 5.13 | 8.46 |
| Vs Observed, km/s | 3451 | 14744 | 2.67 | 0.01 | 0.75 | 1.53 | 2.29 | 2.39 | 2.52 | 4.90 |
| Vs Predicted, km/s | 18195 | 0 | 2.64 | 0.01 | 0.77 | 1.61 | 2.17 | 2.42 | 2.64 | 4.84 |

| Fold-2 | Statistic | | | | | | | | | |
|---|---|---|---|---|---|---|---|---|---|---|
| Feature | N | N* | Mean | SE Mean | StDev | Minimum | Q1 | Median | Q3 | Maximum |
| Temperature Observed, C | 3857 | 14337 | 237 | 6.50 | 403.6 | 16.7 | 63.9 | 70.2 | 100.7 | 1301 |
| Temperature Predicted, C | 18195 | 0 | 260 | 3.16 | 426.7 | 15.2 | 63.0 | 74.0 | 100.8 | 1294 |
| Density Observed, kg/m3 | 212 | 17872 | 2675 | 5.28 | 76.8 | 2467 | 2631 | 2658 | 2736 | 2857 |
| Density Predicted, kg/m3 | 18195 | 0 | 2650 | 0.86 | 116.3 | 2314 | 2628 | 2659 | 2718 | 3012 |
| Log Resisitivity Observed, ohm-m | 356 | 17839 | 2.53 | 0.06 | 1.09 | 0.00 | 1.65 | 2.46 | 3.42 | 5.46 |
| Log Resisitivity Predicted, ohm-m | 18168 | 27 | 2.65 | 0.01 | 1.14 | -0.68 | 2.17 | 2.55 | 3.04 | 7.64 |
| Vp Observed, km/s | 3397 | 14794 | 4.91 | 0.02 | 1.27 | 2.56 | 4.03 | 4.42 | 5.25 | 8.55 |
| Vp Predicted, km/s | 18195 | 0 | 4.79 | 0.01 | 1.30 | 2.67 | 4.04 | 4.33 | 5.13 | 8.46 |
| Vs Observed, km/s | 3393 | 14798 | 2.69 | 0.01 | 0.77 | 1.53 | 2.30 | 2.39 | 2.54 | 4.89 |
| Vs Predicted, km/s | 18195 | 0 | 2.64 | 0.01 | 0.77 | 1.61 | 2.17 | 2.42 | 2.64 | 4.84 |

| Fold-3 | Statistic | | | | | | | | | |
|---|---|---|---|---|---|---|---|---|---|---|
| Feature | N | N* | Mean | SE Mean | StDev | Minimum | Q1 | Median | Q3 | Maximum |
| Temperature Observed, C | 3957 | 14238 | 228.5 | 6.27 | 394.7 | 17.2 | 63.5 | 69.9 | 98.6 | 1301 |
| Temperature Predicted, C | 18195 | 0 | 205.8 | 2.76 | 372.5 | 16.7 | 62.5 | 69.1 | 98.9 | 1299 |
| Density Observed, kg/m3 | 211 | 17984 | 2683 | 5.24 | 76.10 | 2495 | 2629 | 2663 | 2749 | 2975 |
| Density Predicted, kg/m3 | 18195 | 0 | 2669 | 0.87 | 117.9 | 2314 | 2641 | 2698 | 2737 | 2998 |
| Log Resisitivity Observed, ohm-m | 399 | 17796 | 2.58 | 0.06 | 1.18 | -0.62 | 1.66 | 2.49 | 3.42 | 6.87 |
| Log Resisitivity Predicted, ohm-m | 18195 | 0 | 2.92 | 0.01 | 1.27 | -0.17 | 2.23 | 2.82 | 3.41 | 7.71 |
| Vp Observed, km/s | 3476 | 14719 | 4.91 | 0.02 | 1.25 | 2.69 | 4.04 | 4.43 | 5.27 | 8.60 |
| Vp Predicted, km/s | 18195 | 0 | 4.80 | 0.01 | 1.27 | 2.58 | 4.07 | 4.38 | 5.34 | 8.48 |
| Vs Observed, km/s | 3472 | 14723 | 2.69 | 0.01 | 0.76 | 1.53 | 2.30 | 2.39 | 2.54 | 4.90 |
| Vs Predicted, km/s | 18195 | 0 | 2.65 | 0.01 | 0.73 | 1.59 | 2.26 | 2.39 | 2.69 | 4.85 |

**Table 6: Summary of independent testing on numeric model features following development of (80/20) split sets.**

| Fold | Model | R-squared |
|---|---|---|
| 1 | Temperature, C | 99.0% |
| | Density, kg/m3 | 100.0% |
| | Resistivity, ohm-m | 83.1% |
| | P-wave velocity, km/s | 99.1% |
| | S-wave velocity, km/s | 99.1% |

| Fold | Model | R-squared |
|---|---|---|
| 2 | Temperature, C | 99.8% |
| | Density, kg/m3 | 100.0% |
| | Resistivity, ohm-m | 66.0% |
| | P-wave velocity, km/s | 99.1% |
| | S-wave velocity, km/s | 95.5% |

| Fold | Model | R-squared |
|---|---|---|
| 3 | Temperature, C | 99.0% |
| | Density, kg/m3 | 100.0% |
| | Resistivity, ohm-m | 98.9% |
| | P-wave velocity, km/s | 99.1% |
| | S-wave velocity, km/s | 99.3% |

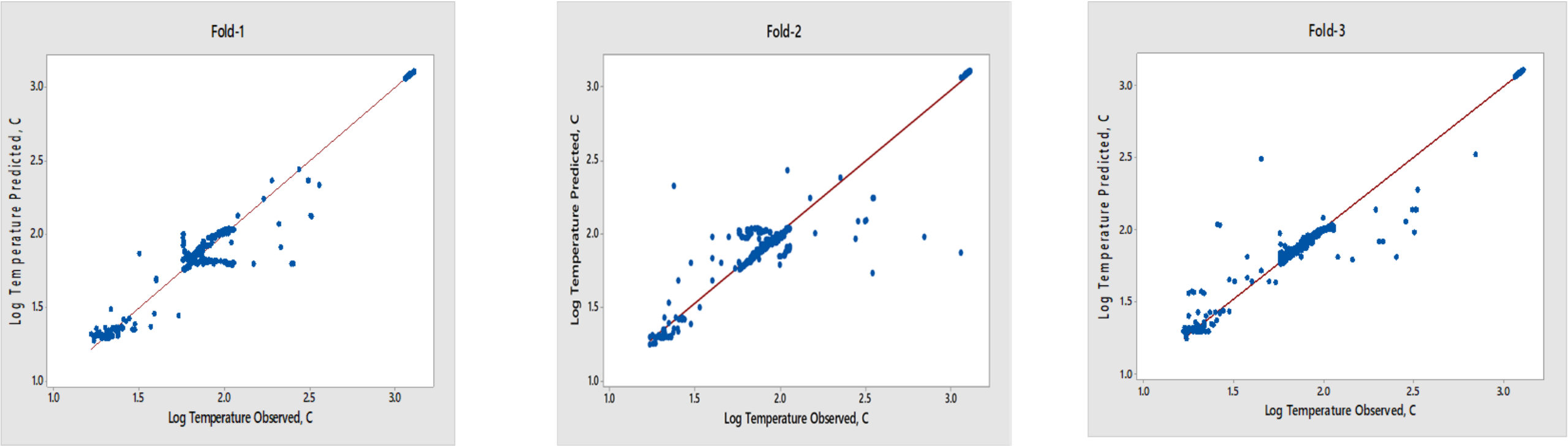

**Figure 18. Cross-plots of temperature with observed (x-axis) and predicted (y-axis) revealing good correspondence for each split set (fold).**

**Table. 7 Kappa statistics (Cohen, 1960) when using machine learning to predict subsurface geologic features. Presence and absence of geologic features based on velocity model of Lahey et al., 1999.**

| Geologic Feature | K-fold | Summary | Statistic | Performance |
|---|---|---|---|---|
| Basalt Layer | 1 | n | 1498 | |
| | | Kappa | 0.441 | Moderate |
| | | Accuracy | 0.650 | |
| | 2 | n | 1498 | |
| | | Kappa | 0.440 | Moderate |
| | | Accuracy | 0.641 | |
| | 3 | n | 1498 | |
| | | Kappa | 0.430 | Moderate |
| | | Accuracy | 0.648 | |

| Geologic Feature | K-fold | Summary | Statistic | Performance |
|---|---|---|---|---|
| Under plating | 1 | n | 1498 | |
| | | Kappa | 0.613 | Substantial |
| | | Accuracy | 0.926 | |
| | 2 | n | 1498 | |
| | | Kappa | 0.664 | Substantial |
| | | Accuracy | 0.928 | |
| | 3 | n | 1498 | |
| | | Kappa | 0.461 | Moderate |
| | | Accuracy | 0.907 | |

| Geologic Feature | K-fold | Summary | Statistic | Performance |
|---|---|---|---|---|
| Oceanic Crust | 1 | n | 1498 | |
| | | Kappa | 0.887 | Almost perfect |
| | | Accuracy | 0.979 | |
| | 2 | n | 1498 | |
| | | Kappa | 0.588 | Moderate |
| | | Accuracy | 0.928 | |
| | 3 | n | 1498 | |
| | | Kappa | 0.869 | Almost perfect |
| | | Accuracy | 0.975 | |

| Geologic Feature | K-fold | Summary | Statistic | Performance |
|---|---|---|---|---|
| Mantle | 1 | n | 1498 | |
| | | Kappa | 0.926 | Almost perfect |
| | | Accuracy | 0.981 | |
| | 2 | n | 1498 | |
| | | Kappa | 0.908 | Almost perfect |
| | | Accuracy | 0.976 | |
| | 3 | n | 1498 | |
| | | Kappa | 0.905 | Almost perfect |
| | | Accuracy | 0.976 | |

Given that the testing results support the ability of MSOM to generalize to independent data, the algorithm is used to simultaneously predict regional numeric and categorical features across the Saddle Rd., Kilauea, and Lower East Rift Zone areas (Figure 19). The set of numeric and categorial feature predictions are presented from a single north looking perspective across the Saddle Rd, Kilauea, and LERZ study areas. The numeric predictions include geophysical features, e.g., temperature, density, resistivity, Vp, and Vp/Vs ratio with the aim of potentially identifying 2D hidden geothermal resources, 2D lithospheric flexure under volcanic loading, and 2D geologic layering.

The temperature predictions reveal a lack of near-surface high temperature geothermal resource below the Saddle Rd profile (Figure 19a). By contrast, temperatures predictions beneath the Kilauea and LERZ areas reveal localized high temperature geothermal anomalies. While anomalous geothermal temperatures were known and exploited in the LERZ, use of the trained model may provide additional geospatial support in their interpretation and ability to identify new drilling prospects with greater resolution. The ability to better target geothermal resources is extended in the next section from these three areas to the whole island using a randomly located set of pseudo-boreholes (characterized by empty cells from the surface to 40km depth) and surface points to develop quasi-3d subsurface image of these same features and various fault systems.

Other geophysical anomalies appear as numeric predictions at the Saddle Rd., Kilauea, and LERZ. For example, there are notable medium-high density anomalies predicted in the right half of the Saddle Rd image from sea level to about 7 km depth (Figure 19b). Other medium-high density anomalies appear at deeper depths in the Kilauea and LERZ areas. By contrast, the Kilauea and LERZ areas reveal large lateral low-density anomalies from sea level to depth of about 7 km and adjacent medium-high density densities to greater depths.

The predicted background resistivity across the study areas is intermediate in value (green in color) with heterogenous high and low resistivity anomalies superimposed from sea level to the model base (Figure 19c). For example, there is one group of medium-low resistivity from sea level to about 3 km depth. A second of medium-low resistivity anomalies crosses the eastern half below of the saddle Rd profile at depths of about 4-8 km, across Kilauea at depths from 4-7 km, and at the western edge of the LERZ at depths of 7-12 km. The bottom of this anomaly is adjacent to a medium-high resistivity anomaly that extends to about 17 km. At this same depth, there is a medium-low resistivity anomaly that extends to about 20 km where the anomaly changes to low-resistivity character. These low-resistivity anomalies represent high conductive features that may be determined to be a particular geothermal process when interpreted in combination with other geophysical anomalies.

The predicted p-wave velocity (Vp) at the study areas ranges from 3.5 to 8.5 km/s (Figure 19d). In general, the upper 1-2 km is of low velocity that progressively increases with depth. The velocity structure of the Kilauea and LERZ are similar characterized loosely as horizontal velocity layers, e.g., <= 4.8 km/s, 6.2 km/s, 7.3 km/s, and >= 8 km/s. Leahy et al. (2010) and Zhong and Watts (2013) identified similar velocity layers referring to them as volcanic, oceanic crust, underplating and mantle. In contrast to the velocity layering at Kilauea and LERZ, the velocity character at Saddle Rd appears to bend downward possibly reflecting downward flexure due to volcanic loading by Mauna Loa and Mauna Kea. Predictions using alternate sets of independent data result in the same or very similar structure across these three study areas.

The predicted background Vp/Vs ratio ranges from 1.5 to 3. Given that Vp/Vs ratios of 3 are often reflective of soil layers, the Vp/Vs ranges in the figure presented (Figure 19e) range from 1.5 to 2 for comparison with other area tomographic studies (Lin et al., 2015). In this figure, the predominant Vp/Vs ratio across the sections is about 1.75 and attributed to competent elastic rock (Gritto, 2022). By contrast, there is a heterogeneous distribution of high and low Vp/Vs ratios appearing laterally adjacent to each other beneath the Saddle Rd and Kilauea areas to depths of about 3 km. At the Saddle Road area, there exists a spatially large high Vp/Vs anomaly of 2.0. This anomaly crosses the western third of the Saddle Rd profile spanning depths from about 3 km to 15 km. Other low Vp/Vs anomalies (1.5-1.6) appear at or near surface locations along profiles of each study area. These near-surface anomalies are interpreted as gaseous fluids, such as discussed at the LERZ (Nugraha, et al., 2016). The low Vp and low Vp/Vs anomalies in the upper 2 or 3 km at Kilauea and the LERZ may be caused by the presence of steam (PGV, 1991) or volatiles (Lin et al., 2015). At much greater depths, such as below about 17 km at Saddle Rd and 23 km at LERZ, where Vp and Vs are high and Vp/Vs anomalies > 1.8, may indicate the presence of partial or molten magma. This is particularly true toward the west and across the first third of the Saddle image.

In the next section, the trained MML model is used to simultaneously predict categorial feature predictions with results presented from the same north looking perspective across the Saddle Rd, Kilauea, and LERZ study areas (Figure 20). The preliminary categorical feature predictions focus on geologic features, such as basalt, (oceanic) crust, underplating, with the aim potentially identifying 2D lithospheric flexure under volcanic loading and 2D geologic layering. From these results, we interpret the boundary between the predicted magmatic underplating and mantle to be the Moho. Based on this interpretation, depth to the Moho under Kilauea and LERZ is about 13 km but under Mauna Loa and Mauna Kea is estimated to vary between 13 km (edge of flexure) to 23 km (bottom of flexure).

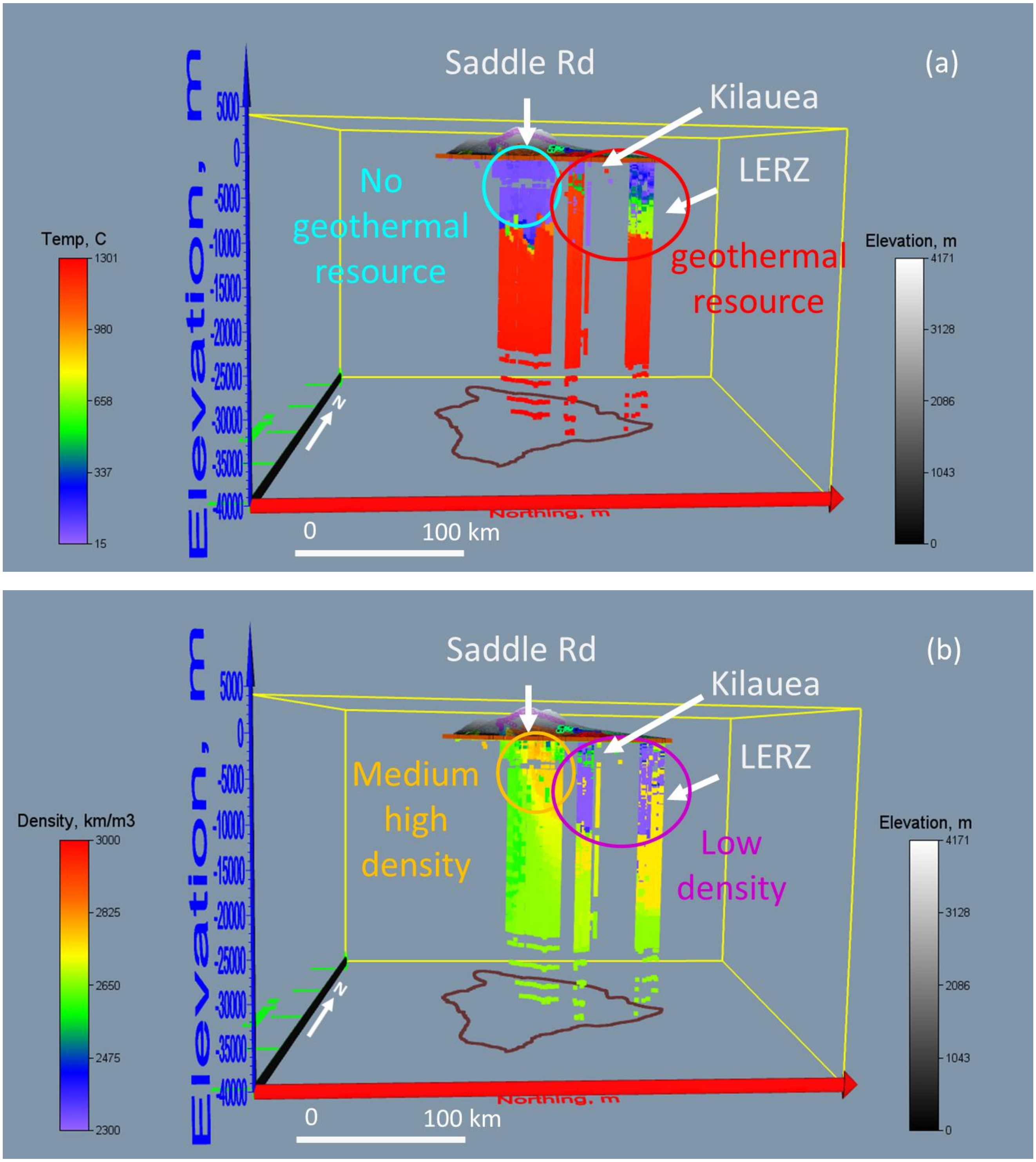

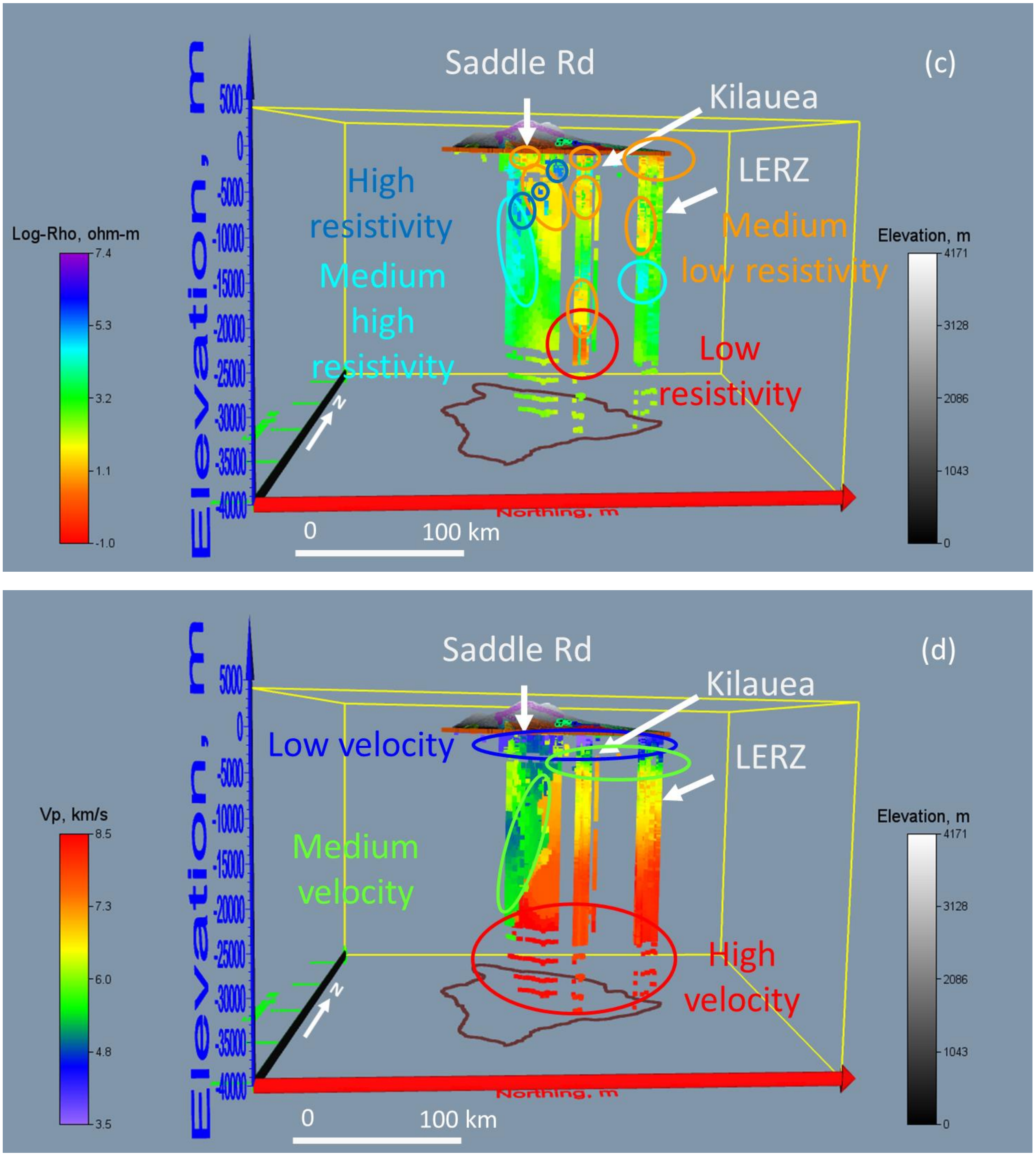

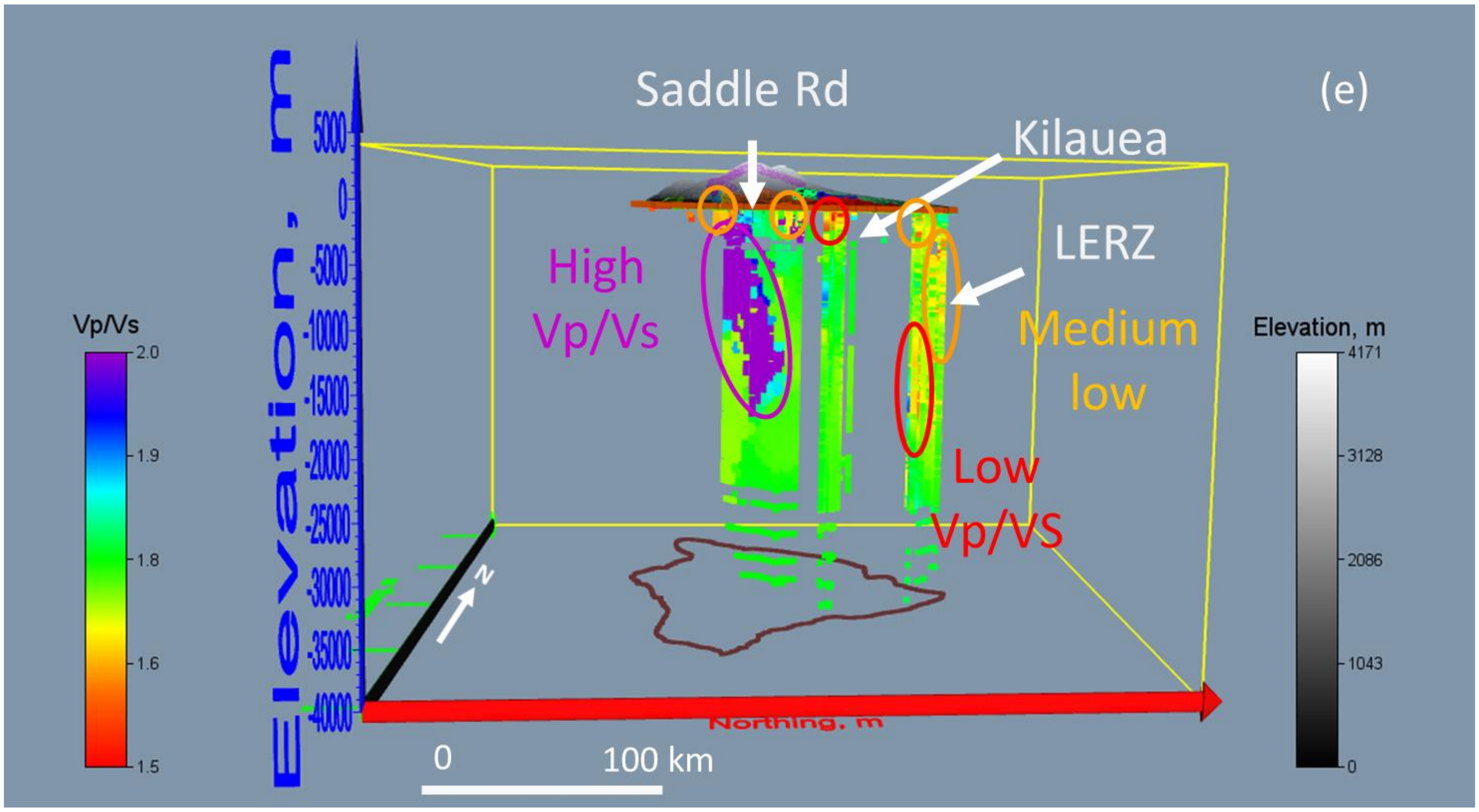

**Figure 19. Regional simultaneous feature predictions across Saddle Rd., Kilauea, and Lower East Rift Zone, using the third independent split set (fold-3): (a) temperature, C; (b) density, kg/m3; (c) log-resistivity (ohm-m), (d) Vp, km/s; and (e ) Vp/Vs ratio. Low and high anomalies are annotated in each of the plots.**

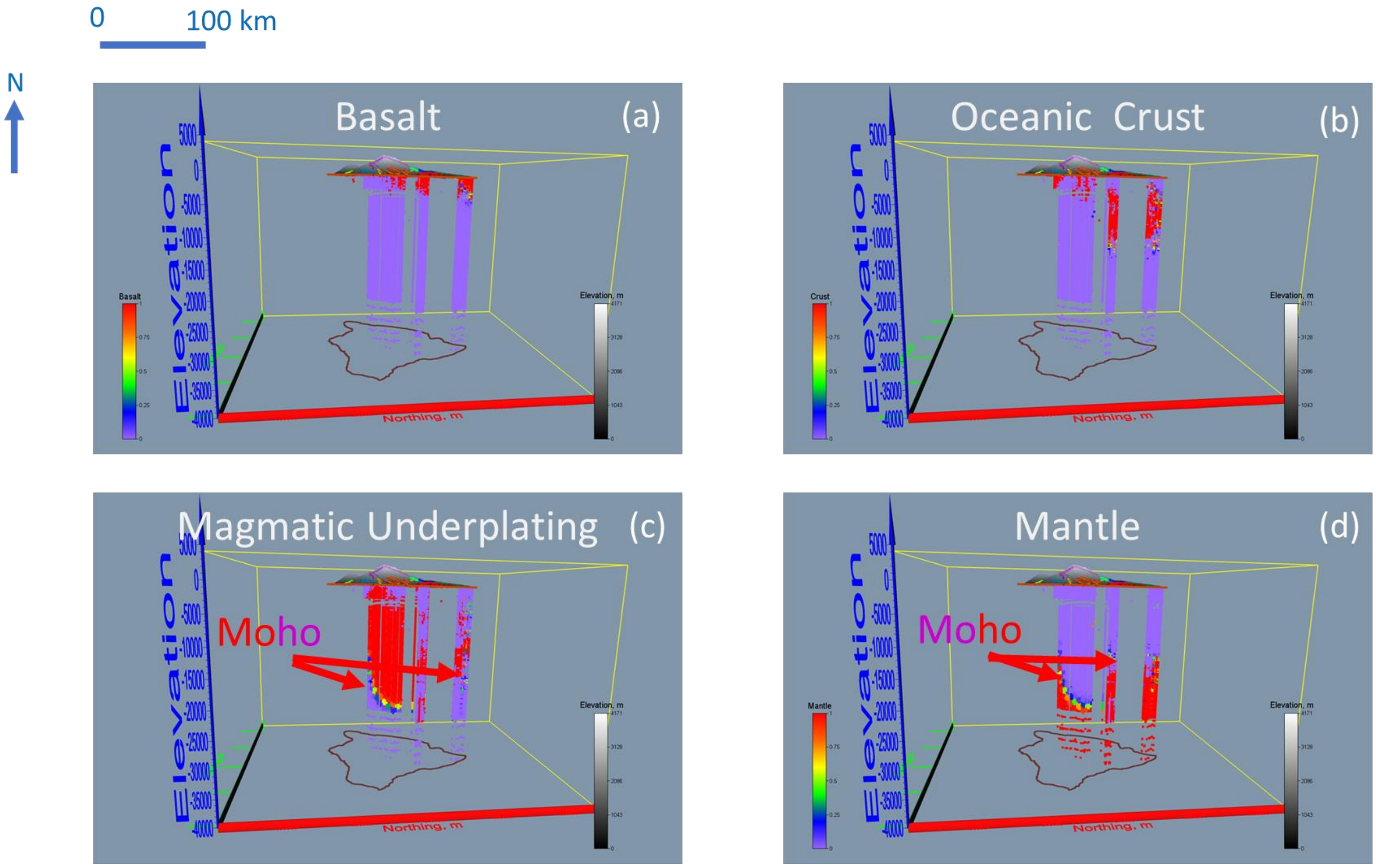

**Figure 20. Simultaneous categorical feature predictions across Saddle Rd., Kilauea, and Lower East Rift Zone, using the first independent split set (fold-1): (a) temperature, C; (b) density, kg/m3; (c) log-resistivity (ohm-m), (d) Vp, km/s; and (e ) Vp/Vs ratio. The interpreted location of the Moho is consistent across the Kilauea and LERZ but appears much deeper beneath the Saddle Rd profile. The deeper locations of the Moho beneath Saddle Rd might be attributed to volcanic loading of Mauna Kea and Mauna Loa.**

4.2.2.3 Island-wide Assimilation and Prediction

In this section, results are presented for the independent island-wide assimilation and prediction of features using the Saddle Rd., Kilauea, and Lower East Rift Zones records together with 2500 randomly located surface and 150 random randomly located pseudo-boreholes (characterized by empty cells from the surface to a depth of 20 km below which a linear temperature gradient is applied; Putrika, 1997) scattered across the island. The aim of this section is to identify 3d hidden geothermal resources, 3d lithospheric flexure under volcanic loading, 3d geologic layering, 3d earthquake – fault association, and 3d fault characterization. The objectives are to: (1) predict 3d geophysical (numeric) features, e.g., temperature, density, resistivity, Vp, Vp/Vs ratio; and (2) predict 3d geologic (categorical) features, e.g velocity layers basalt, crust, underplating, and mantle. The feature predictions are presented as a single independent set (fold-3) in plan-view together with (a) eight prominent island fault systems (b), regional feature predictions, (c) island-wide feature predictions below sea level, and (d) island-wide feature predictions below and above sea level. The fault systems and rifts on Hawai'i are numbered and color coded; for example, fault systems: Hilina=1 (red), Kahuku =2 (red-orange), Kaoiki = 3 (orange), Kealakekua = 4 (dark green), Koae = 6 (dark blue), Kohala = 7 (light blue); and rifts: Kilauea = 5 (light green) and Mauna Loa = 8 (purple). The inclusion of regional feature predictions in each figure provides a means to visually evaluate the continuity among features when extending the regional application to the whole island using randomly placed pseudo-boreholes and randomly placed near surface stations. Given that the phase2 and phase3 data sets will be improved by application of 3d and joint inversions, the results presented herein are considered preliminary with the focus on identifying anomalies with minimal interpretation.

In the first set of figures, independent temperature predictions are presented over the range of 10-1300 C (Figure 21) and over the range of 10- >350 C (Figure 22). In both figures, the continuity of temperature predictions at (b) the regional and (c) below sea-level island-wide temperature suggests that the extension of feature prediction using the pseudo-boreholes may be useful in future studies. Also, several temperature anomalies are present with reviewing these figures. Those temperature anomalies at the LERZ area are known and being produced for their geothermal resource, whereas the geothermal anomalies beneath the Kilauea caldera and area southeast of the caldera correspond with recent and ongoing eruptions of molten lava. In addition to the LERZ and Kilauea temperature anomalies, there are two anomalies on the east and west flanks of Mauna Kea and one southwest of Mauna Loa near to the Mauna Loa fault system.

Independent density predictions are presented over the range of 2300-3000 kg/m3 in figure 23. Following the application of the island-wide prediction process, the prominent low-density feature (2300 kg/m3) in the LERZ area expands southwest parallel to the coast and north of the Hilina Fault system. Other relatively high-density anomalies (2800 kg/m3) are evident at and above sea level beneath Mauna Kea and Mauna Loa. The largest density anomaly (3000 km/gm3) appears near the surface and directly below the Mauna Kea caldera.

There are several log-resistivity anomalies are identified across the island. The most prominent low resistivity (high conductivity) features appear in the upper 3 km at the LERZ and Kilauea areas (Figure 24). Another medium-low resistivity anomaly occurs along the western end of Saddle Rd. Other low resistivity features are localized south of the Saddle Rd at the northwest and southwest flanks of Mauna Loa both west of Mauna Loa Fault system. Various spatially large medium to medium high resistivity features is apparent in the island-wide prediction sets. The largest high resistivity (purple) feature on the big island is associated with the upper Kaoiki fault system. This feature appears in the above sea level island-wide predictions. The second largest medium high resistivity (dark blue) also appears only in the above sea level predictions and along the Mauna Loa Fault. The presence of both these features in the above sea level prediction sets may suggest that the associated Kaoiki and Mauna Loa Fault systems may be relatively shallow.

The prediction of Vp is successfully extended from the regional study areas across the big island (Figure 25). The velocity predictions presented in panels b-d reveal a spatially variable gradient as described by others when studying the region from Kilauea southeast toward the coast (Park et al., 2007; Lin et al.,2014). The implementation of the island-wide prediction process resulted in increasing the anomalous velocity below the Saddle Rd from 3.5 km/s to about 6 km/s. The revised upward Vp prediction is more reflective of oceanic crust which seems plausible considering published discussions on big island volcanic loading and downward flexure of the oceanic crust reaching a minimum between the Mauna Kea and Mauna Loa volcanoes as described by others (Leahy et al., 2010, Zhong and Watts, 2013, Klein, 2016). Further supporting this hypothesis is the change from layer-cake velocity structure beneath Kilauea and LERZ to a downward and compressed velocity trend reaching a minimum point south of the Saddle Rd and west of Mauna Loa fault. West of this region there are no pseudo-boreholes and therefore Vp predictions to define the likelihood of structural symmetry with possible rising flexural limb reflective of the layer cake east of Mauna Loa Fault.

The prediction of Vp/Vs ratio is successfully extended from the regional study areas across the big island (Figure 26). Multiple high and low Vp/Vs anomalies exist (1.5-2.0) that can be combined with Vp and low log-resistivity anomalies to better understand spatial geothermal resources on the big island. For example, in the shallow (upper 2 or 3 kms) vicinity of the LERZ and Kilauea areas the anomalously low Vp/Vs ratio and correspondingly low log-resistivity might be interpreted as presence of gaseous fluids. The presence of steam in the HPGA and other geothermal wells of the LERZ are examples of this behavior (Puna Geothermal Venture, 1991). A similar anomalous region in the vicinity of the Koae Fault zone can be interpreted likewise. By contrast, the higher is the Vp/Vs ratio, the softer is the geologic material. In this case, the high Vp and Vp/Vs ratios together with low log-resistivity may reflect a partial melt. One example is at lower depths beneath the LERZ, where the combination of high Vp, high Vp/Vs and low log-resistivity might be associated with the upward flow of partially molten magma. Another deeper region of anomalously high Vp, high Vp/Vs ratio, and low log-resistivity exists at depts below about 17 km beneath Saddle Rd. and southward under Mauna Loa might be interpreted as the location of partially molten magma.

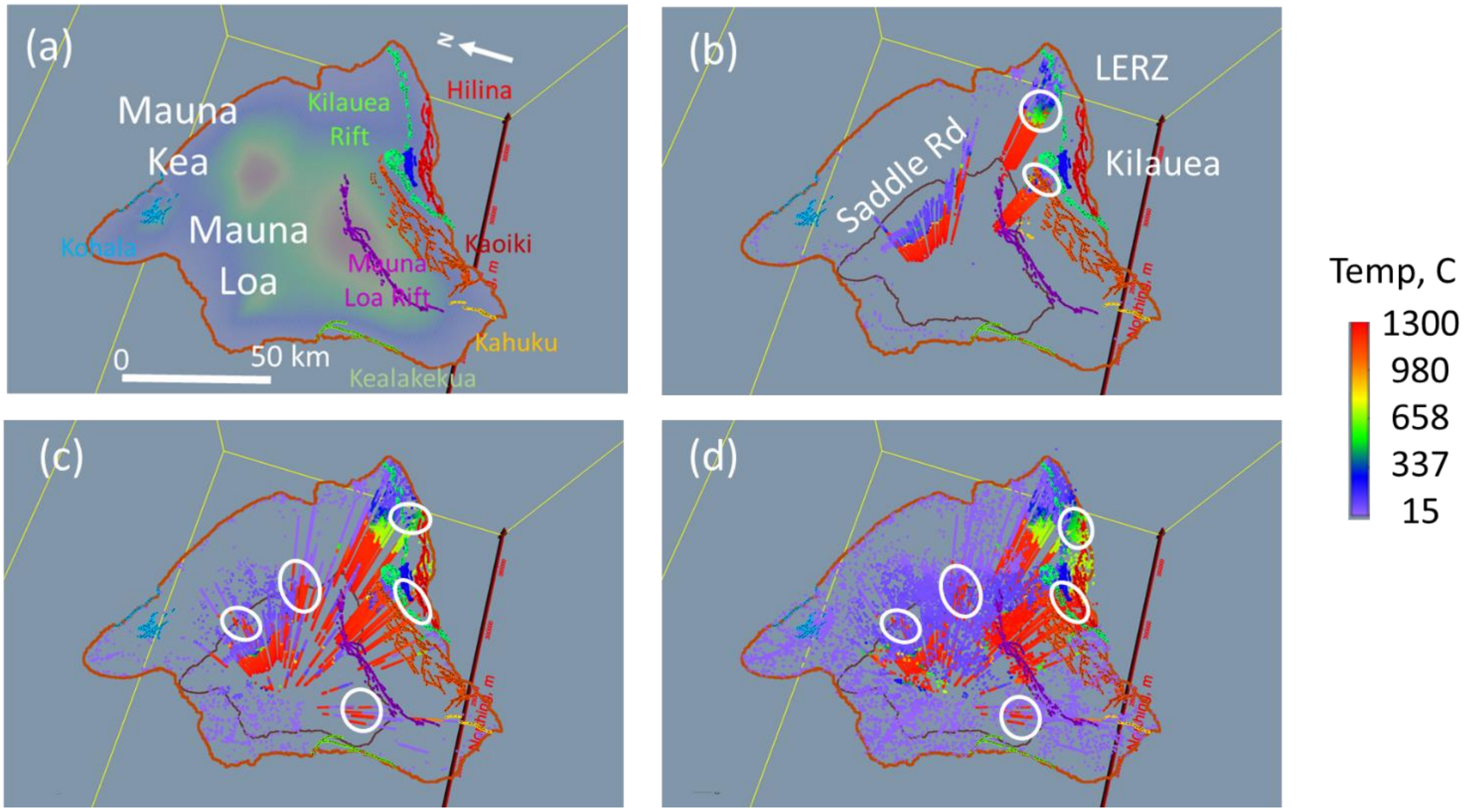

**Figure 21. Comparison of temperature (10-1300 C) predictions at the regional and island scales: (a) fault systems: Hilina=1 (red), Kahuku =2 (red-orange), Kaoiki = 3 (orange), Kealakekua = 4 (dark green), Koae = 6 (dark blue), Kohala = 7 (light blue), and rifts: Kilauea = 5 (light green), Mauna Loa = 8 (purple); (b) independent (fold-3) regional below sea level temperature predictions across Saddle Rd, Kilauea and Lower East Rift Zone areas, (c) independent island-wide below sea level temperature predictions based on assimilation of regional features and 100 random pseudo-boreholes, (d) independent island-wide above and below sea level temperature predictions based on assimilation of regional features and 100 random pseudo-boreholes and 2500 random surface locations. Geothermal anomalies are identified by white circles.**

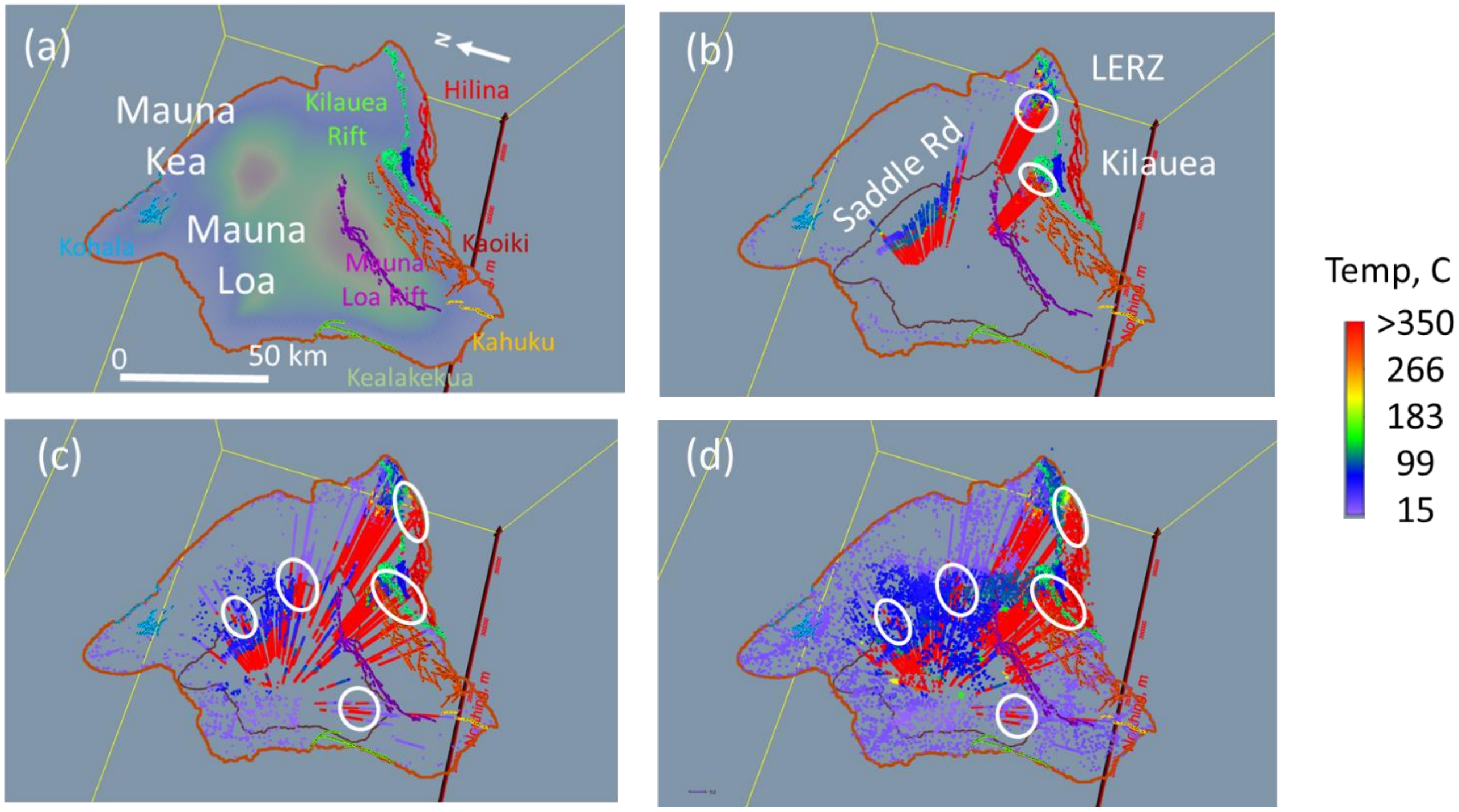

**Figure 22. Comparison of temperature (10->350 C) predictions at the regional and island scales: (a) fault systems: Hilina=1 (red), Kahuku =2 (red-orange), Kaoiki = 3 (orange), Kealakekua = 4 (dark green), Koae = 6 (dark blue), Kohala = 7 (light blue), and rifts: Kilauea = 5 (light green) and Mauna Loa = 8 (purple); (b) independent (fold-3) regional below sea level temperature predictions across Saddle Rd, Kilauea, and Lower East Rift Zone areas, (c) independent island-wide below sea level temperature predictions based on assimilation of regional features and 100 random pseudo-boreholes, (d) independent island-wide above and below sea level temperature predictions based on assimilation of regional features and 100 random pseudo-boreholes and 2500 random surface locations. Hot colors represent high temperature areas and cool colors represent low temperature areas.**

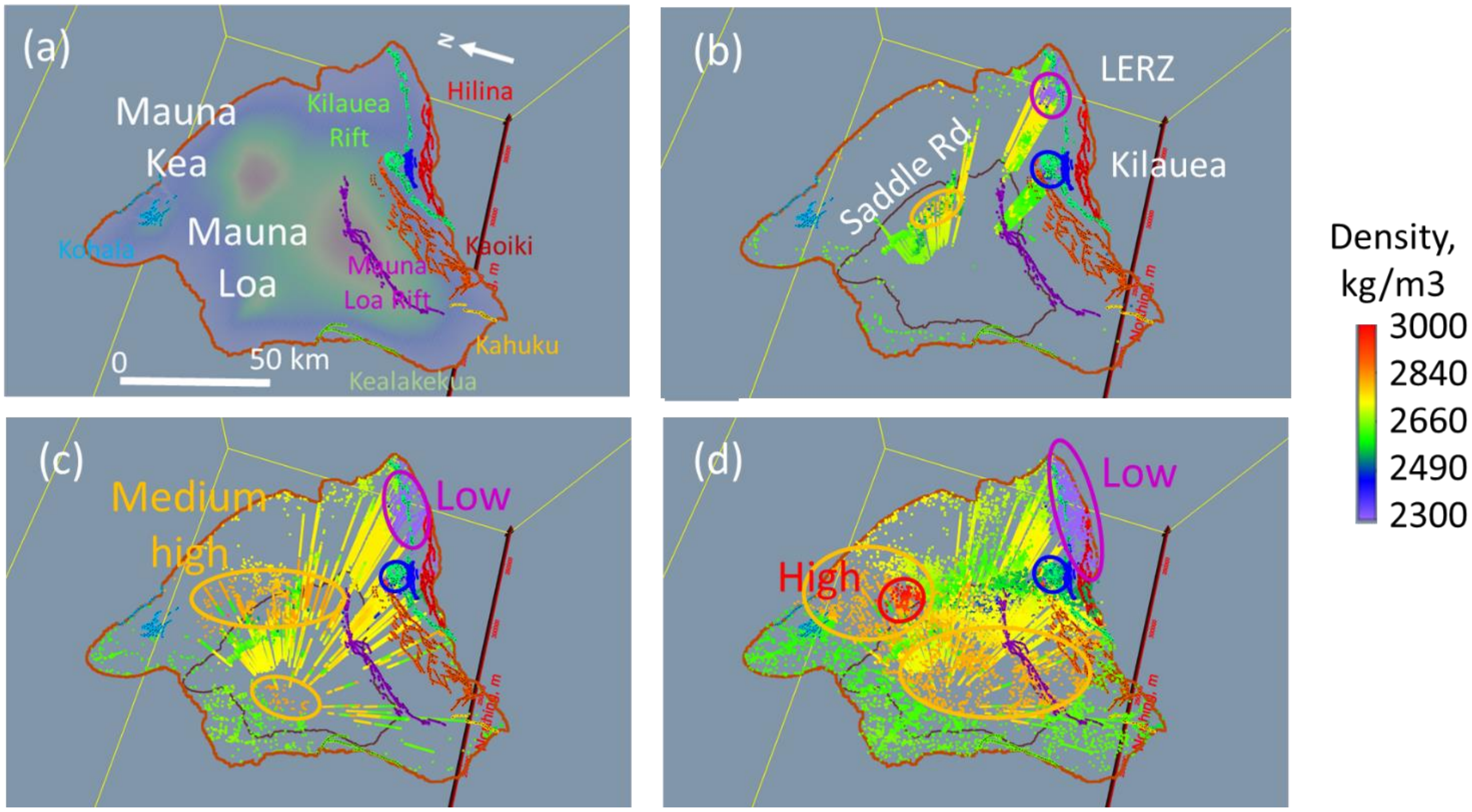

**Figure 23. Comparison of density (2300-3000 kg/m3) predictions at the regional and island scales: (a) fault systems: Hilina=1 (red), Kahuku =2 (red-orange), Kaoiki = 3 (orange), Kealakekua = 4 (dark green), Koae = 6 (dark blue), Kohala = 7 (light blue), and rifts: Kilauea = 5 (light green) and Mauna Loa = 8 (purple); (b) independent (fold-3) regional below sea level density predictions across Saddle Rd, Kilauea, and Lower East Rift Zone areas, (c) independent island-wide below sea level density predictions based on assimilation of regional features and 100 random pseudo-boreholes, (d) independent island-wide above and below sea level density predictions based on assimilation of regional features and 100 random pseudo-boreholes and 2500 random surface locations. Hot colors represent high density areas and cool colors represent low density areas.**

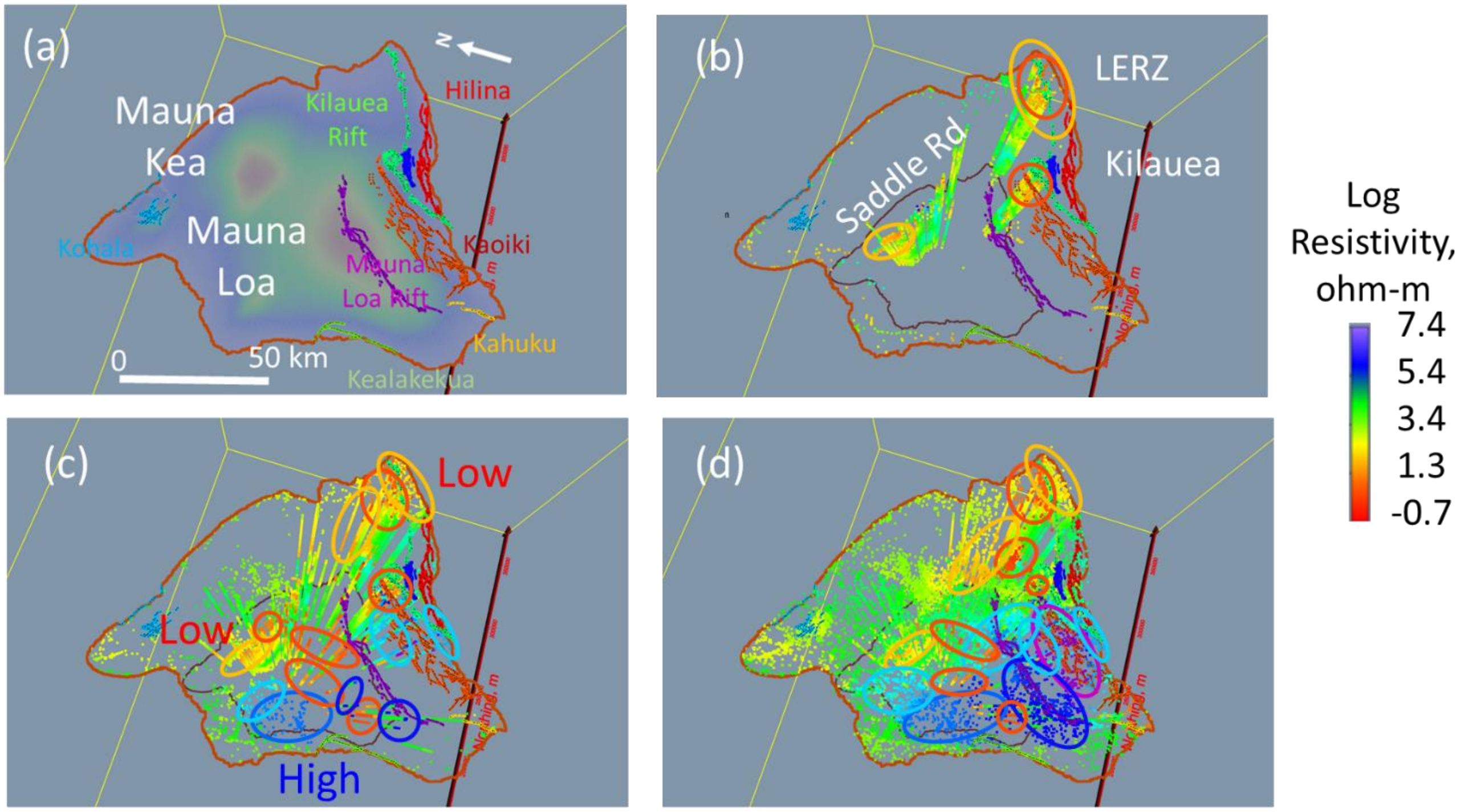

**Figure 24. Comparison of Log resistivity (-0.7 to 7.7 ohm-m) predictions at the regional and island scales (a) fault systems: Hilina=1 (red), Kahuku =2 (red-orange), Kaoiki = 3 (orange), Kealakekua = 4 (dark green), Koae = 6 (dark blue), Kohala = 7 (light blue), and rifts: Kilauea = 5 (light green) and Mauna Loa = 8 (purple); (b) independent (fold-3) regional below sea level log-resistivity predictions across Saddle Rd, Kilauea, and Lower East Rift Zone areas, (c) independent island-wide below sea level log-resistivity predictions based on assimilation of regional features and 100 random pseudo-boreholes, (d) independent island-wide above and below sea level log-resistivity predictions based on assimilation of regional features and 100 random pseudo-boreholes and 2500 random surface locations. Hot colors represent low resistivity (high conductivity) areas and cool colors represent high resistivity low conductivity) areas.**

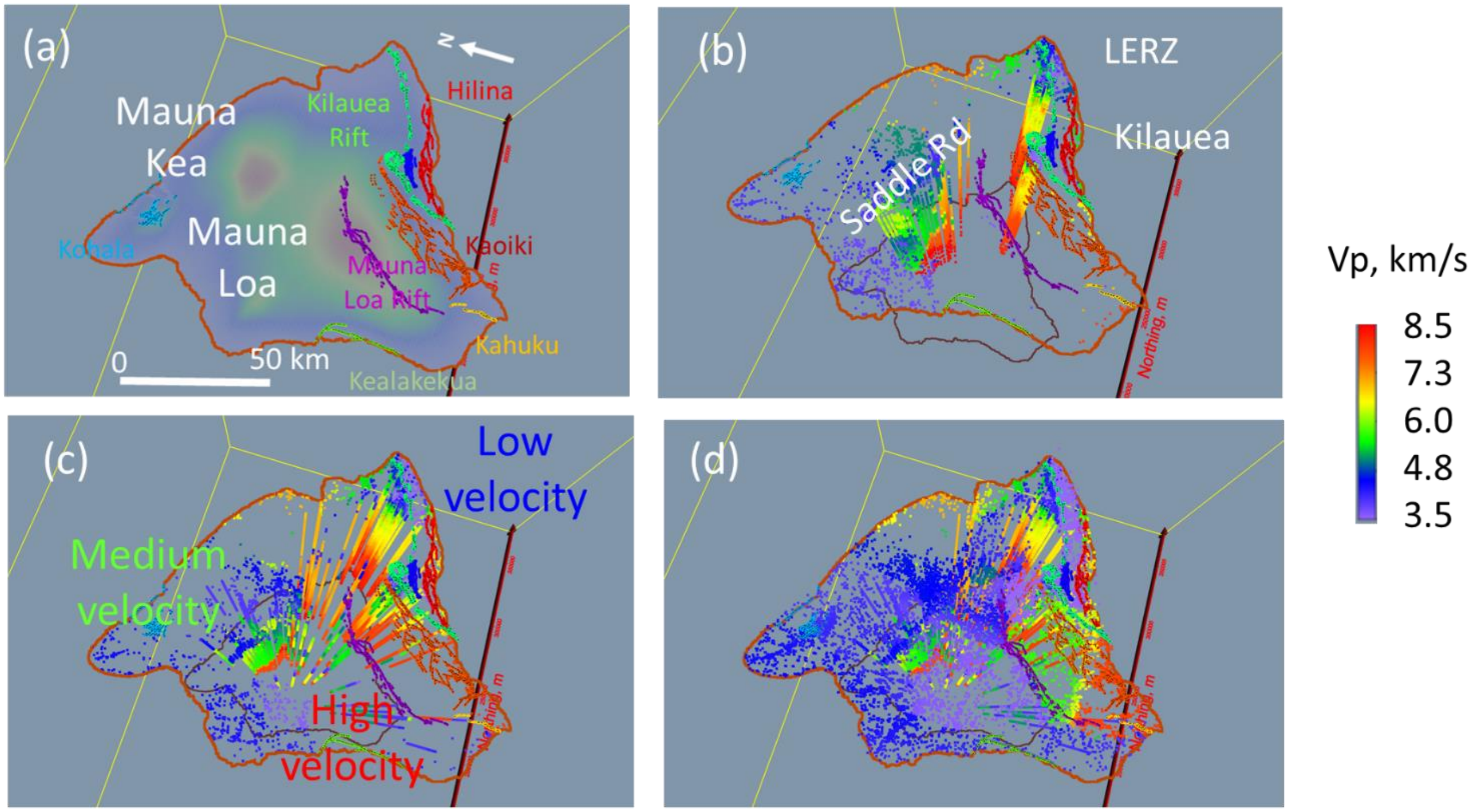

**Figure 25. Comparison of Vp (3.5 to 8.5 km/s) predictions at the regional and island scales: (a) fault systems: Hilina=1 (red), Kahuku =2 (red-orange), Kaoiki = 3 (orange), Kealakekua = 4 (dark green), Koae = 6 (dark blue), Kohala = 7 (light blue), and rifts: Kilauea = 5 (light green) and Mauna Loa = 8 (purple); (b) independent (fold-3) regional below sea level Vp predictions across Saddle Rd, Kilauea, and Lower East Rift Zone areas, (c) independent island-wide below sea level Vp predictions based on assimilation of regional features and 100 random pseudo-boreholes, (d) independent island-wide above and below sea level Vp predictions based on assimilation of regional features and 100 random pseudo-boreholes and 2500 random surface locations. Hot colors represent high values of Vp and cool colors represent low values of Vp ratio. The velocity predictions in the panel (b-d) reveal a spatially variable gradient described by others when studying the region from Kilauea southeast toward the coast (Park et al., 2007; Lin et al.,2014).**

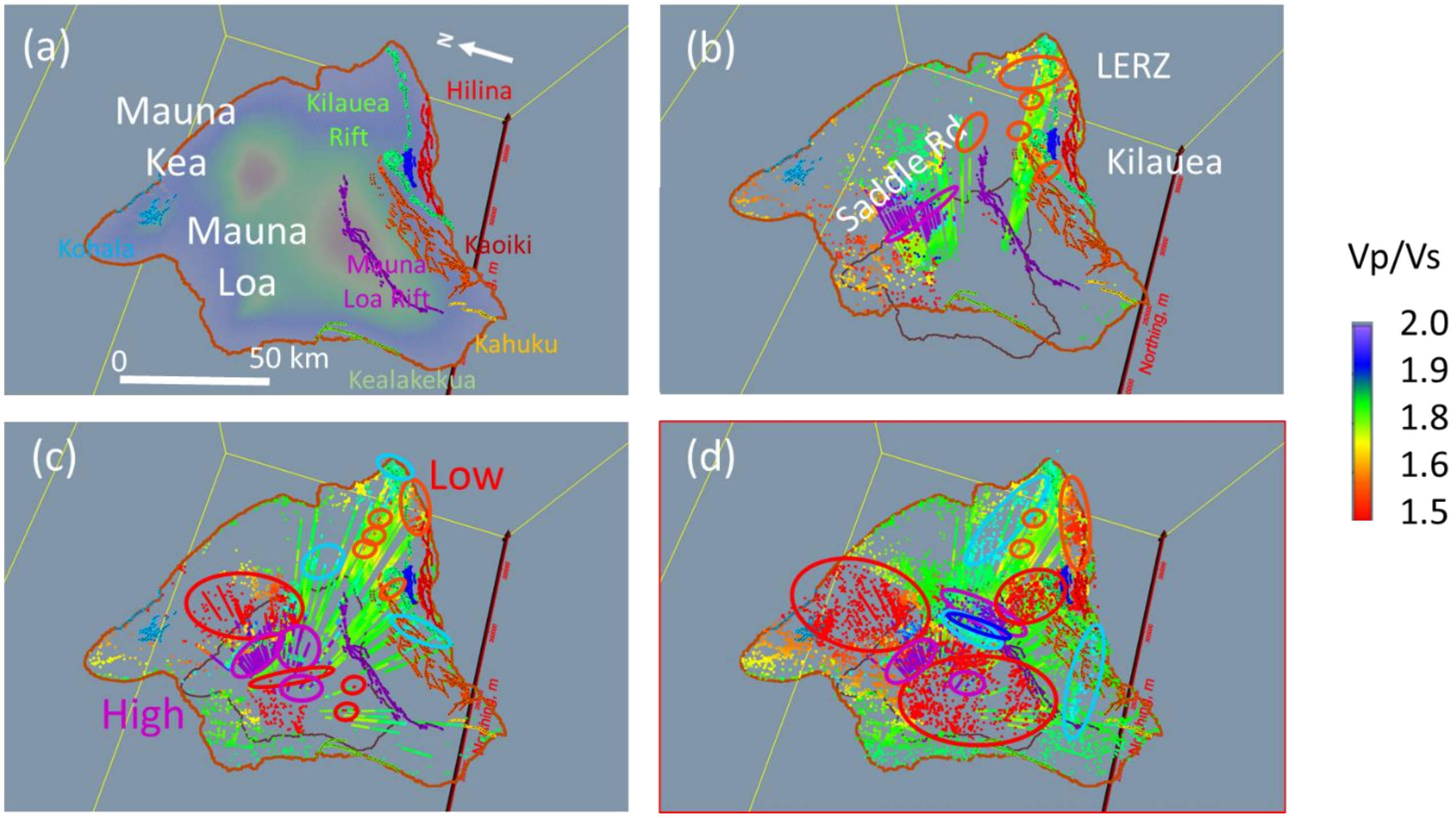

**Figure 26. Comparison of Vp/Vs ratios (1.5 to 2.0) predictions at the regional and island scales: (a) fault systems: Hilina=1 (red), Kahuku =2 (red-orange), Kaoiki = 3 (orange), Kealakekua = 4 (dark green), Koae = 6 (dark blue), Kohala = 7 (light blue), and rifts: Kilauea = 5 (light green) and Mauna Loa = 8 (purple); (b) independent (fold-3) regional below sea level Vp/Vs predictions across Saddle Rd, Kilauea, and Lower East Rift Zone areas, (c) independent island-wide below sea level Vp/Vs predictions based on assimilation of regional features and 100 random pseudo-boreholes, (d) independent island-wide above and below sea level Vp/Vs predictions based on assimilation of regional features and 100 random pseudo-boreholes and 2500 random surface locations. Hot colors represent low values of Vp/Vs ratio and cool colors represent large values of Vp/Vs ratio.**

The predictions of quasi-3d geologic layering using categorical constraints based on known and predicted velocity are presented in Figure 27. The simultaneous prediction of basalt, oceanic crust, magmatic underplating and mantle, and layers appear in the additional panels (b-d). The upper right and lower left panels show different viewing perspectives of the predicted magmatic underplating beneath the island of Hawai’i. These figures further support the notion of oceanic flexure under the Hawai’ian volcanic load (Leahy et al., 2010, Zhong and Watts, 2013, Klein, 2016).

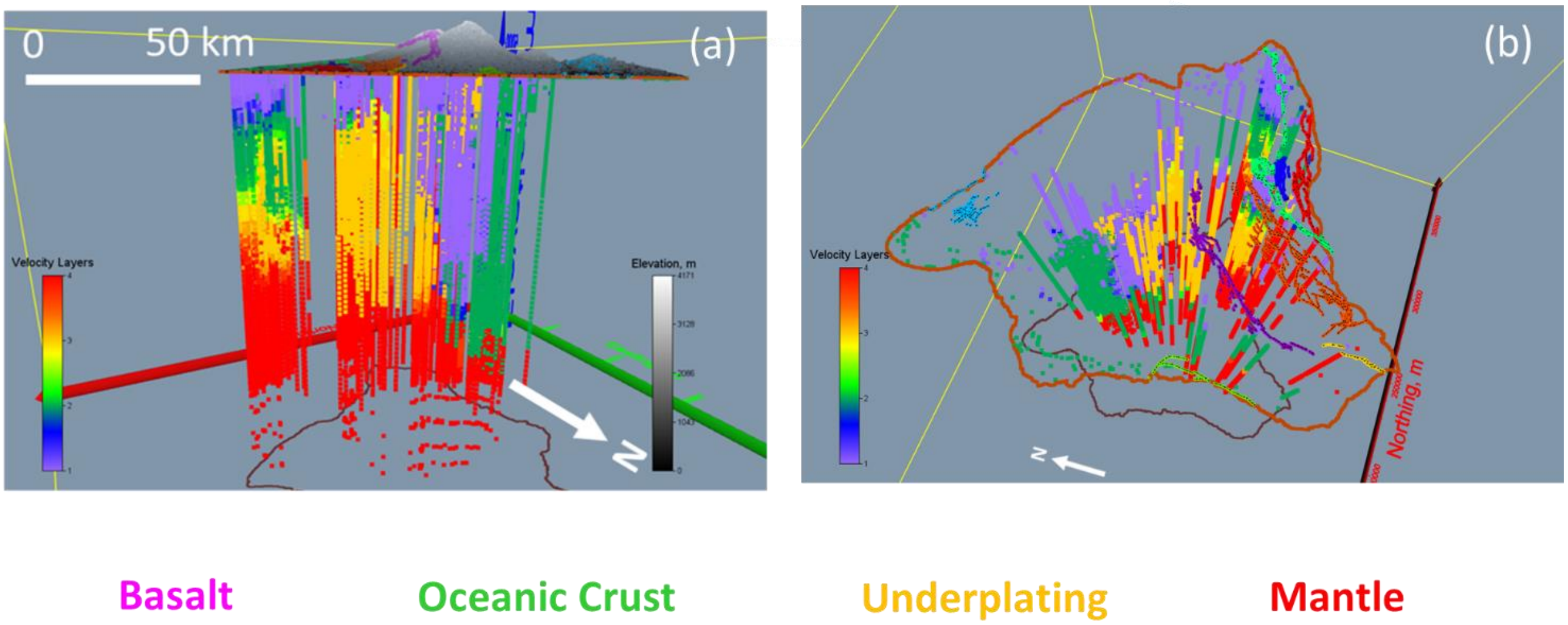

**Figure. 27. This figure reveals predicted 3d geologic layering using categorical constraints based known and predicted velocity. The velocity predictions in the upper left panel reveal a spatially variable gradient described by others when studying the region from Kilauea toward the coast. The simultaneous estimation of all features including Vp, basalt, oceanic crust, magmatic underplating and mantle, and layers appear in the additional panels. The upper right and lower left panels show different viewing perspectives of the predicted underplating beneath Hawai’i. These figures support the common notion of oceanic flexure under the Hawai’ian volcanic load (Leahy et al., 2010, Zhong and Watts, 2013, Klein, 2016).**

The predicted 3d earthquake-fault system associations presented in Figure 28. Some of the prominent features include the association of earthquakes with distinct fault systems and rifts. Note that the earthquake and fault systems are predicted simultaneously with their association indicated by the same color. For example, there are earthquake clusters associated with the Hilina fault system 1 (red), the Kaoiki fault system number 3 (colored), the Kilauea rift number 5 (light green), and the Koae fault system number 6 (blue). The most prominent shallow earthquakes appear to be associated with the Kilauea rift system, the intermediate deep earthquakes appear to be associated with the Hilina fault system, whereas the deepest earthquakes appear to be associated with the Kaoiki fault system.

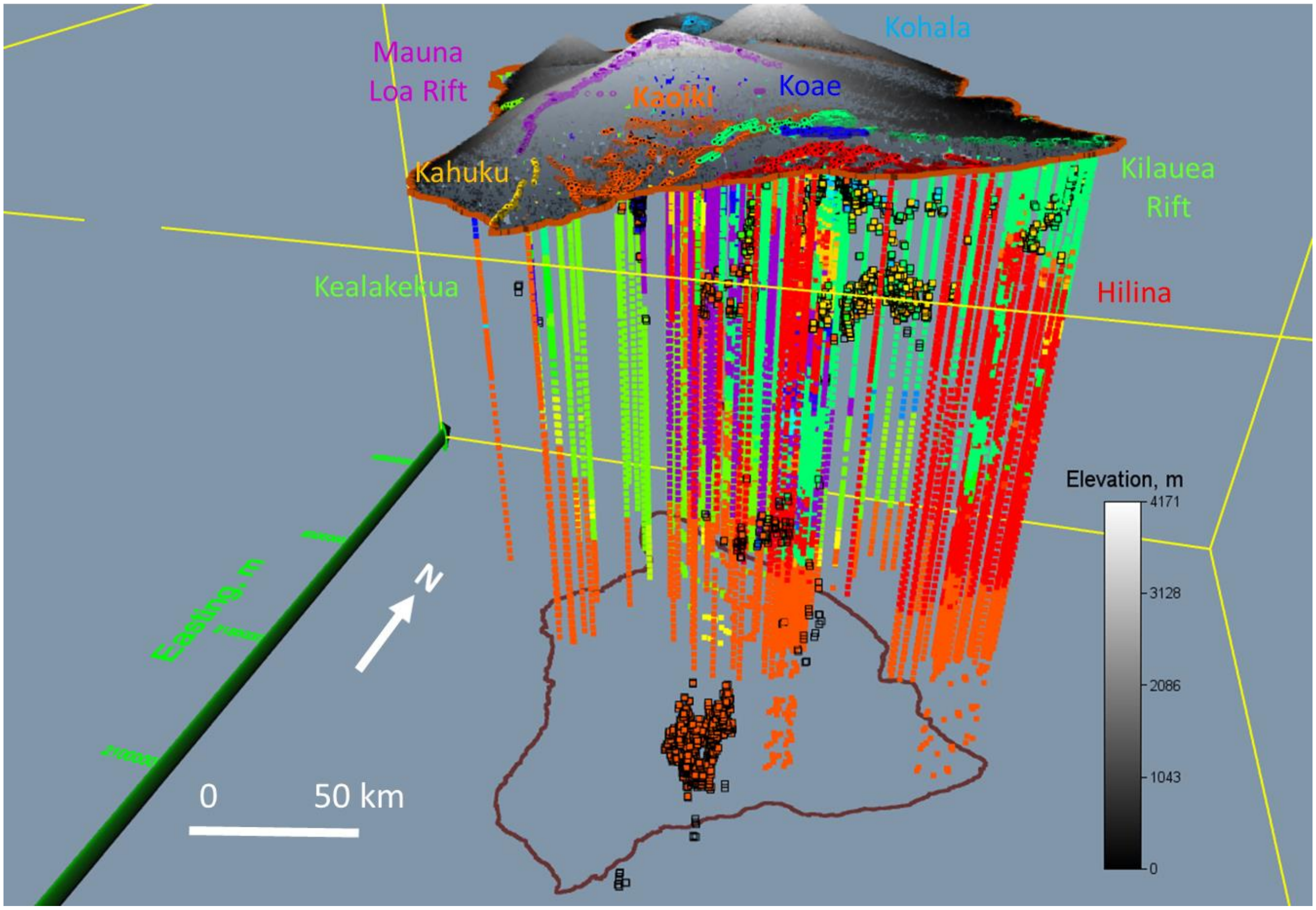

**Figure. 28. 3d Fault – earthquake characterization predictions for the eight fault systems from different perspectives. Some of the prominent features include the association of earthquakes with distinct fault systems.**

Predicted characteristics of selected faults on the big island are presented in Figure 29. The two left panels (a and c) provide top views (below sea level) that reveal an approximately NE trending Kaoiki fault system and approximately northward tending Kahuku fault system. The right two panels (b and d) are side looking northwest views toward the Kilauea - LERZ coastline. These figures reveal modeled fault aspects including the depths and trends of the Koae and Hilina fault systems. For this case, the Hilina fault appears to start mid-way along the coast trending northeast in the direction of the Puna ridge, whereas the Koae fault appears to start toward maintaining a shallow profile trending toward to the southwest.

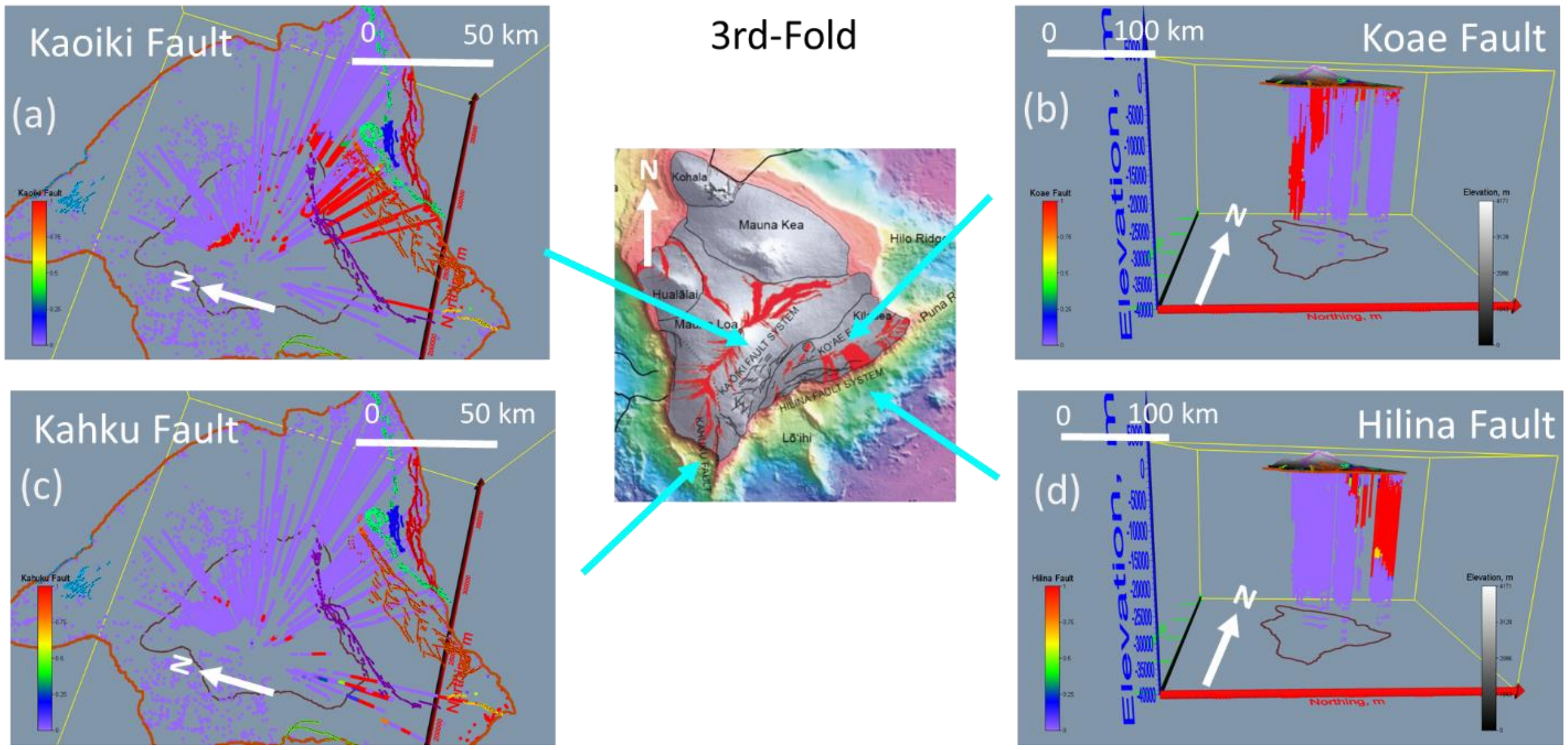

**Figure. 29. Predicted characteristics of selected faults on the big island. The two left panels show top views that reveal an approximately NE trending Kaoiki fault system and approximately northward tending Kahuku fault system. The right two panels are looking northwest toward the Kilauea - LERZ coastline. These figures reveal modeled aspects including the depths and trends of the Koae and Hilina fault systems. For this case, the Hilina fault appears to start mid-way along the coast trending northeast in the direction of the Puna ridge, whereas the Koae fault appears to maintain a shallow profile trending toward the southwest.**

## 5. SUMMARY

In this study, a multimodal machine learning (MML) workflow was proposed, features selected, and modified self-organizing map (MSOM) algorithm trained and tested for predicting groundwater and geothermal feature characteristics at the island of Lānaʻi, and geothermal characteristics at the island of Hawaiʻi. Despite each case study using different subsets of the Hawaiʻi Play Fairway hydrogeologic, geothermal, and geophysical data, there associated models both generalized to independent test data (randomly shuffled data sets split with 80% for training and 20% for testing) despite being characterized as sparse, spatially limited with different support volumes, and uncertain.

In the first case study, the trained Lānaʻi model was used to predict continuous 3d features from which interesting merges of clusters characterized five 3d Geothermal Stratigraphic Units. The predicted features include state variables (such as hydraulic head, temperature, chloride concentration), physical properties (such as density, resistivity, specific capacity – permeability surrogate), geology (basalt and dike – generalized for plutonic and dike material), aqueous chemistry (major ions, metals, isotopes), and aqueous properties (pH, specific conductance, dissolved oxygen). Inspection of the predicted features revealed possible geologic attributes, such as basalt, dike swarms, sill, pluton, batholith, and Moho; possible groundwater attributes, such as downward groundwater recharge, saltwater intrusion, heterogenous chloride concentration, such as zones of freshwater, brackish water, and saline water; and geothermal attributes, such as upward convective hydrothermal transport along a gradient from the very hot Moho (758 C) to the near surface heterogeneous temperature distribution ranging from warm (65 C) to cool (20 C). Based on this information, two freshwater resources were identified from the surface to below 1 km depth: a warm water (50-65 C) region with intermediate specific capacity, low chloride concentration (<50 mg/l), and a cool water (about 20C) region with low specific capacity, low chloride concentration (<300 mg/l). Two geothermal resources were identified: one in the vicinity of the caldera where there is a warm (65 C) water at or near to the surface warmer down to 1 km depth, and a very hot hydrothermal plume being transported upward from the Moho (about 750 C) to commercially viable drilling depths of about 2.0 km to 6 km depth with temperatures of about 107 C to 275 C. The ability to determine Geothermal Stratigraphic Units (GSUs) from independent set of hydrogeologic, geothermal and geophysical measurements support their potential use in simultaneous characterization of combined groundwater and geothermal resources. Further, this novel approach affords the possibility for direct assignment of GSUs and their features to numerical model cells (or nodes), thereby minimizing ambiguity in the conceptualization to numerical modeling process including initial starting parameter values, boundary conditions, and geostatistical constraints in support of the calibration process. The MML workflow techniques used herein are recognized as novel in this application and their initial success warrants further research. The performance metrics in the Lānaʻi case study provide encouragement to continue this line of groundwater-geothermal research as part of the new Island Heat project funded by DOE's GTO.

In the second case study, the trained Hawai'i model was first used to simultaneously predict regional numeric and categorical features across the Saddle Rd., Kilauea, and Lower East Rift Zone areas. The set of numeric and categorial predictions included geophysical features, e.g., temperature, density, resistivity, Vp, and Vp/Vs ratio from which preliminary interpretations revealed 2D hidden geothermal resources, 2D lithospheric flexure under volcanic loading, and 2D geologic layering appear identified. Next, the regional model was extended to island-wide by introducing 2500 random surface and 150 pseudo-boreholes randomly scattered across the island to predict 3d geophysical (numeric) features, e.g., temperature, density, resistivity, Vp, Vp/Vs ratio; and 3d geologic (categorical) features, e.g velocity layers basalt, crust, underplating, and mantle were undertaken facilitating the potential identification of 3d hidden geothermal resources, 3d lithospheric flexure under volcanic loading, 3d geologic layering, 3d earthquake – fault association, and 3d fault characterization. Despite success in achieving the aim and objectives of this case study, the model can be improved in ways suggested as next steps in phase 2 and phase 3.

## ACKNOWLEDGEMENTS

This work was supported by a generous donation from the Pulama Lānaʻi to Lautze through the University of Hawai'i Foundation, and the U.S. Department of Energy's Geothermal Technologies Office grant under award numbers DE-EE0006729 and WBS3.1.1.15.